\documentclass[%
bibnotes,
 amsmath,amssymb,
]{revtex4}

\pdfoutput=1

\usepackage{graphicx}
\usepackage{dcolumn}
\usepackage{bm}
\usepackage{float}
\usepackage{color}
\usepackage{hyperref}
\usepackage{mathrsfs}
\usepackage{rotating}
\usepackage{epstopdf}

\newcommand{\be}{\begin{eqnarray}}
\newcommand{\ee}{\end{eqnarray}}

\begin{document}

\title{Higgs couplings and new signals from Flavon-Higgs mixing effects\\ within 
multi-scalar models}
\author{J. Lorenzo Diaz-Cruz,\footnote{\texttt{jldiaz@fcfm.buap.mx}} \hspace{0.1 cm}  Ulises J. Salda\~na-Salazar\footnote{\texttt{ulisesjesus@protonmail.ch}}}
\affiliation{ C. A. Nueva F\'isica en Aceleradores y el Cosmos, \\
Facultad de Ciencias F\'isico-Matem\'aticas,\\
Benem\'erita Universidad Aut\'onoma de Puebla, C.P. 72570, 
Puebla, Pue., M\'exico.}


\begin{abstract}
Testing the  properties of the Higgs particle discovered at the LHC  and searching 
for new physics signals, are some of the most important tasks of Particle Physics today. 
Current measurements of the Higgs couplings to fermions and gauge bosons, seem 
consistent with the Standard Model, and when taken as a function of the particle mass, 
should lay on  a single line. However, in models with an extended Higgs sector
the diagonal Higgs couplings to up-quarks, down-quarks and charged leptons, could lay on 
different lines, while non-diagonal  flavor-violating Higgs couplings could appear too. 
We describe these possibilities within the context of multi-Higgs doublet models 
that employ the Froggatt-Nielsen (FN) mechanism  
to generate the  Yukawa hierarchies. Furthermore,  one of the doublets 
can be chosen to be of the inert type, which provides a viable dark matter candidate.
The mixing of the Higgs doublets with the flavon field, can provide plenty of
interesting signals, including:
i) small corrections  to the couplings of the SM-like Higgs, 
ii) exotic signals from the flavon fields,
iii) new signatures from the heavy Higgs bosons.
These aspects are studied within  a specific model with 3+1 Higgs doublets and a singlet FN field. 
Constraints on the model are derived from the study of K and D mixing and the Higgs search at the LHC. For last, the implications from the latter aforementioned constraints to the FCNC top decay $t\to ch$  are presented too.  

\end{abstract}

\maketitle

\section{\label{sec:Intro} Introduction}
Particle Physics is facing an exciting time thanks to the
discovery of a Higgs-like particle with $m_h=125-126$ GeV at the 
LHC \cite{higgs-atlas:2012gk,higgs-cms:2012gu}, which  has provided 
a definite test of the mechanism of electroweak symmetry breaking \cite{Gunion:1989we}.  
Furthermore, the Higgs mass  value agrees quite  well with the range predicted by  
electroweak precision tests \cite{Erler:2007sc}, which confirms the success of the 
Standard Model (SM). 
Current measurements of its spin, parity, and couplings, seem also consistent with 
the SM. The fact that the LHC has verified the SM linear realization of spontaneous symmetry 
breaking (SSB),  could also be taken as an
indication that Nature likes scalars, which was in doubt years ago. 

On the other hand, several scenarios for physics beyond the SM have been conjectured 
in order to address some of its open problems, such as hierarchy, flavor, unification, 
etc \cite{Pomarol:2012sb}. 
However, the LHC has not detected, so far, any sign of new physics, but has provided bounds on 
its scale ($\Lambda$), which are now entering into the multi-TeV range.
This is bringing some discomfort, and casting some doubts about
the theoretical motivations for those new physics scenarios with a mass scale 
of order TeV. This is particularly disturbing for  the concept of naturalness, 
and its supersymmetric implementation, since the bounds 
on the mass of superpartners are passing the TeV limit too \cite{Martin:1997ns,Kane:2006hd}.
 However, one has to wait for this new LHC run, which has higher energy and luminosity, 
 to collect enough data
in order to have stronger limits, both on the search for new particles, such as heavier
Higgs bosons \cite{Arganda:2013ve,Arganda:2012qp,Chakraborty:2013si}, and for
precision tests of the SM Higgs properties.

In particular, it will be very important to study the Higgs couplings at the LHC and 
future colliders,  in order to test the Higgs identity. Being the remnant of SSB, 
the Higgs particle couples to a pair of massive gauge bosons, with an strength proportional 
to its mass. As the Higgs doublet induces the fermion masses too, its couplings
with fermions is also proportional to the fermion mass; these are
flavor-conserving (FC) couplings. 
However,  so far the LHC has tested only a few of these couplings,
i.e. the ones with the heaviest SM fermions and $W$ and $Z$ bosons. 
Then, some questions arise immediately:

\begin{itemize}
 \item Do the masses of all fermion types (up-, down-quarks and leptons) arise from a
       single Higgs doublet? or, are there more Higgs multiplets participating in the game?
       \cite{Botella:2016krk}
 \item Are the Higgs couplings to fermions diagonal in flavor space? or, is it possible to have 
 flavor-violating Higgs couplings?
 \item Is there any hope to measure the Higgs couplings with the lightest quarks and leptons?
 \item Is there a connection between Higgs physics and the Dark Matter (DM) problem? \cite{DiazCruz:2007be}
 \item What are the implications of the Higgs properties for physics in the far ultraviolet? \cite{Veltman:1980mj}
\end{itemize}

Non-standard Higgs couplings, including the flavor violating (FV) ones, are 
predicted in many models of physics beyond the SM, for instance 
in the general 2HDM  \cite{Grossman:1994jb,Branco:2011iw, DiazCruz:2010yq,DiazCruz:2004tr}.  
The 2HDM is a particular realization of SM extensions that
contain additional scalar fields, which can have non-aligned 
couplings to the SM fermions and through its mixing with the  Higgs boson,
new FV Higgs interactions can be induced at tree-level.  Mixing of SM fermions 
with exotic ones, could also  induce FV Higgs couplings at tree-level \cite{Cotti:2002zq}.
 It could also happen that the Higgs sector has  diagonal couplings to the SM fermions at tree-level, 
but the presence of new particles with a non-aligned flavor structure, which couple both to the Higgs 
and to the SM fermions, will induce corrections to the diagonal Yukawa couplings and/or FV Higgs couplings at 
loop levels. This is realized for instance in supersymmetry, where the sfermion/gauginos can  have 
non-diagonal couplings to Higgs bosons and SM fermions \cite{DiazCruz:2002er}.The Higgs boson with a non-minimal flavor structure was called 
 {\it{a more flavored Higgs boson}}  in our earlier work \cite{DiazCruz:2002er}.

After the LHC delivered the Higgs signal, many papers have been devoted to study
non-standard Higgs couplings, and the constraints on deviations from the SM   
\cite{Espinosa:2012ir, Espinosa:2012im, Klute:2012pu, Cheung:2013rva, Ellis:2014dva,Grinstein:2013fia}. 
 Although such deviations could be discussed within an effective  {Lagrangian} approach,
it is also important to discuss them within an specific model, where such deviations 
could be interpreted and given a context. 

In this paper, we shall explore the implications for the Higgs sector within multi-Higgs models 
that also address some aspects of the flavor problem. Namely, we will work within the context of 
a class of models that employ Abelian flavor symmetries and the Froggatt-Nielsen mechanism 
(with a singlet flavon field), to generate the  Yukawa hierarchies. Furthermore,  in these models 
one of the Higgs doublets  is chosen to be of the inert type, which provides a viable dark matter candidate.
In general, the scalar spectrum will include a light SM-like Higgs boson, as well as
extra heavy bosons, which can mix with the flavon fields. The properties of
 each of these states will show some peculiar manifestations from their interaction with
the flavor sector, which in turn could induce plenty of interesting signals that could 
be searched at future colliders.  For instance, the diagonal couplings of the light SM-like Higgs boson 
could receive small corrections, while FV couplings could be induced at small rates too. 
It is known that within the SM,  the Higgs couplings  to fermions and gauge bosons, 
as function of the particle mass, lay on a single straight line. 
However, we will show that in our model, fermion couplings could lay on distinct lines.
{The logic is the following. Within the two Higgs doublets model of type I, where all fermion masses come from a single doublet, the Higgs couplings to fermions will also lay on a single straight line. On the other hand, in the type II case, where up-type quarks get masses from one Higgs while
down-type quarks and charged leptons get masses from a second Higgs, the Higgs couplings to fermions will lay on two different straight lines. Therefore, to get Higgs-fermion couplings laying on three different straight lines we need to use a distinct Higgs doublet to generate masses for each fermion type.}
Larger rates are predicted  within our model for the FV couplings of the heavy Higgs bosons
and even more striking signals are expected for the flavon fields.

All these aspects are studied in detail for a specific model 
with 3+1 Higgs doublets, which realizes both features: the diagonal Higgs-fermion couplings
lay on multiple lines and its magnitude deviates from the SM predictions, while at the same time
allows for the presence of non-diagonal couplings.
The organization of our paper goes as follows. We present first (in Section \ref{Sec:II}), 
the motivation for considering muti-Higgs models with a flavon field. We also discuss 
some general features of the possible mass spectrum, as well as the signals that are expected for 
the light and heavy Higgs bosons of this type of models.
The Yukawa Lagrangian of the specific 3+1 model, with Higgs masses and couplings is studied in Section \ref{Sec:III}.
In Section \ref{Sec:IV}, we discuss the constraints on the parameters of the model, and identify some
scenarios that could be further studied at the LHC. 
Then,  the implications of FV Higgs couplings are presented in Section \ref{Sec:V}, 
focusing on the  top  quark FCNC transitions  $t\to ch$ and the possible contributions to the K and D mixing. 
Concluding remarks are included in Section \ref{Sec:VI}, including a 
brief discussion about the  search for the flavon particle itself,
and the possible ways to test the Higgs couplings with the lightest fermions.

\section{Higgs and flavon spectrum  in Multi-Higgs model with Froggatt-Nielsen mechanism}
\label{Sec:II}

One of the open problems of the SM is the proliferation of parameters, in particular
the quark and lepton masses,  the angles and phases appearing in the Cabibbo-Kobayashi-Maskawa (CKM) and Pontecorvo-Maki-Nakagawa-Sakata (PMNS) mixing
matrices. The measured values for these parameters show a hierarchical pattern in the quark
sector and close to maximal values in the leptonic one. A variety of ideas have been proposed
in the last 20-30 years trying to obtain a deeper understanding of this problem, with some
degree of  success (for a review see for instance \cite{Fritzsch:1999ee} or for more up to date one see \cite{King:2013eh}). These range from a phenomenological approach 
(textures) to GUT-inspired relations, flavor symmetries and radiative generation, to mention a few of them. Recently, nonetheless, a phenomenological approach using solely the 
hierarchy in the fermion masses, $m_3 \gg m_2 \gg m_1$, provided for the first time a generical
treatment wherein the mixing patterns for both quarks and leptons can be simultaneously understood \cite{Hollik:2014jda}. 

Within the flavor symmetry approach, the Froggatt-Nielsen mechanism offers a partial 
explanation of the Yukawa hierarchies. Specifically, one assumes that above some scale $M_F$, 
there is a flavor symmetry  that forbids the  appearance of the Yukawa couplings; 
SM fermions are charged under this symmetry (taken to be of the Abelian type $U(1)_F$).
However, the Yukawa matrices can arise  through  non-renormalizable operators that are included 
in an effective Lagrangian of the type: 
\begin{equation}
 {\cal{L}}^Y_{eff} =  \alpha^a_{ij} (\frac{ S }{M_F} )^{n_{ij}} \bar{F}_i f_j  \tilde{\Phi}_a + h.c.
\end{equation}
which involves the left-handed fermion doublet $F_i$,  right-handed
fermion singlets $f_j$ ($i,j=1,2,3$), and the Higgs doublets ${\Phi}_a$ ($a=1,..N$).
The abelian charges of these fields add to $n_{ij}$, while the flavon field $S$
is assumed to have flavor charge equal to -1, such that ${\cal{L}}_{eff}$ is $U(1)_F$-invariant.
Then,  Yukawa matrices arise after the spontaneous breaking  of the flavor symmetry,
and its entries are expressed in terms of the ratio of the
flavon vacuum expectation  value ($<S>$) and the heavy mass 
scale ($M_F$). i.e. $\lambda_x = (\frac{<S>}{M_F})^{n_x}$. 
The  heavy mass scale $M_F$ represents the mass of heavy fields that  
transmit such symmetry breaking to the quarks and leptons, thereby 
generating the Yukawa matrices.

In this paper, we study the implications of the mixing of such flavon fields with the 
Higgs sector, which becomes even more relevant after the LHC measurements of the Higgs couplings 
with gauge bosons and fermions are reaching a precision era.
For the flavon field, we consider the minimal setting, that is, we employ an Abelian 
flavor symmetry that is broken by the vev of a  complex scalar flavon field, singlet under
the SM gauge symmetry.  There are several possible Abelian charges for the SM fermions, which produce 
realistic Yukawa matrices, and here the most salient features will be discussed. 
The number of Higgs doublets is also left opened in this section, in order to discuss the expected 
phenomena that will appear from the scalar spectrum.

\section{Higgs couplings within a 3+1 Higgs model}
\label{Sec:III}

We consider for definitiveness a model with four Higgs 
doublets which are denoted as $\Phi_1, \Phi_2, \Phi_3,$ and  $\Phi_4$.
Three of them are responsible for each giving masses to a fermion type, 
i.e. $\Phi_1$ gives masses to the up-type quarks, while 
$\Phi_2$ and $\Phi_3$ to the d-type quarks
and leptons, respectively. This model implies that the Higgs couplings could deviate from the SM
predictions. The Higgs couplings to gauge bosons deviate 
from the SM, because the model includes several doublets developing a vev, 
while the diagonal Yukawa couplings show deviations because fermion masses 
are generated by a different Higgs doublet. At this point, having a Higgs per fermion type,
flavor changing neutral currents (FCNC) are not a worry. Indeed, 
natural flavor conservation
emerges, as Glashow and Weinberg proved it  \cite{Glashow:1976nt}, whenever all fermions of a given electric charge do not couple to more than one Higgs. 

{For the sake of completeness, we include a DM candidate. Its sole main purpose is to provide a complete model with a rich phenomenology to be studied through different steps in future studies. Thus, the} remaining 
doublet ($\Phi_4$) is used as a DM candidate via the imposition of a discrete symmetry.
The addition of this symmetry is made in such a way that this doublet 
is of the inert-type, and therefore the lightest particle within this doublet is stable,
thus becoming a viable (scalar) DM candidate \cite{Ma:2006fn}. {Further details about which symmetry and the correct assignment of its charges to the fields is here omitted as it is out of the scope of this work and has no effect in the study of the Higgs-fermion couplings.}
We shall only discuss  briefly the effects of  Dark Matter on the Higgs sector.
Although the detailed properties of the Dark Sector will depend
on the specific form of the Higgs potential, the essential information will be contained in 
the mass of the DM candidate and its coupling with the SM-like Higgs boson, 
which will allow to calculate the invisible
Higgs decay and compare with the LHC limits on the invisible Higgs branching ratio.
However, given that a  model with 2+1 Higgs doublets has already been discussed 
in ref. \cite{Grzadkowski:2010au},
which includes extra sources for CP violation and has a rich phenomenology, which
also improves the inert dark matter model \cite{Krawczyk:2013jta}, 
it seems reasonable to expect that
our model could also satisfy the DM contraints;
the study of the DM implications is left for future work.

Furthermore, we also include a SM singlet, $S$, which participates
in the generation of the Yukawa hierarchies, \`a la Froggatt-Nielsen, and thus has FV couplings. {The FN mechanism is known to be able to generate the Yukawa hierarchies, and
	from a low energy point of view it can be treated as an effective field theory, which just needs to reproduce the Yukawa pattern. Thus, from this perspective, we do not need to specify the details of the model regarding whether the flavor symmetry is discrete or continous, global or local. There are several interesting constraints on the models by looking at the UV completion, for instance, to study gauge coupling unification. Anomaly cancellation is another issue that provides constraints on the fermion spectrum. However, we want to work on the low-energy constraints, i.e. ${\cal O}$(TeV) effects, and thus decided to leave for future work the detailed aspects of the model that are dependent on the UV completion. }

This singlet mixes with the Higgs doublets, and induces FV Higgs couplings. 
So we pay a fare cost: our natural flavor conservation is lost but in return the hierarchy
between the fermion masses is theoretically included.
The case with mixing of the SM Higgs doublet with a flavon singlet, and its
low energy phenomenology was studied in ref. \cite{Dorsner:2002wi}, while a detailed study of
high-energy aspects of the model were studied in \cite{Tsumura:2009yf}, while a recent study of flavon-Higgs mixing 
phenomenology appeared in \cite{Berger:2014gga}.
Besides serving us as a specific model to test the pattern of Higgs couplings,
and new physics, our 3+1 model can be discussed within the context of models
where  some flavor symmetry is linked with the DM stability \cite{Hirsch:2010ru}. 
More recently, the possibility that a model with a single Higgs doublet and a FN singlet, could 
help to explain the 750 GeV resonance hinted at the LHC was explored in 
\cite{Bolanos:2016aik}.

\subsection{Yukawa Lagrangian for a multi-Higgs model with Flavon-Higgs mixing}

We start with the Yukawa Lagrangian, \`a la Froggatt-Nielsen, given by
\begin{equation}
- {\cal{L}}_Y =    \rho^u_{ij}  ( \frac{ S }{\Lambda_F} )^{n_{ij}} \bar{Q}_i d_j  \tilde{\Phi}_1
                + \rho^d_{ij}  (\frac{ S }{\Lambda_F})^{p_{ij}} \bar{Q}_i u_j \Phi_2  
                + \rho^l_{ij}  (\frac{ S }{\Lambda_F})^{q_{ij}} \bar{L}_i l_j \Phi_3  +h.c. \; ,
\end{equation}
where $n,p,q$ denote the charges of each fermion type under some unspecified Abelian flavor 
symmetry, which will help to explain the fermion mass hierarchy. The field $S$ 
is a complex flavon field, while $\Lambda_F$ denotes the flavor scale  
and $\rho^f_{ij}$ ($f=u,d,l$) are some $O(1)$ coefficients. 
The Higgs doublets are written as: $\Phi_i = (\Phi^+_i,  \Phi^0_i)^T$.

We expand to linear order in the flavon terms,
\begin{eqnarray}
	S = \frac{1}{\sqrt{2}} (u + s_1 + i s_2),
\end{eqnarray}
as follows,
\begin{equation}
 ( \frac{S}{\Lambda_F})^{n_{ij}} =  \lambda^{n_{ij}}_F \left[1+ \frac{n_{ij}}{u} (s_1+i s_2) \right],
\end{equation}
where 
\begin{eqnarray}
	\lambda_F= \frac{u}{\sqrt{2}\Lambda_F} \simeq 0.22,
\end{eqnarray}
 is of the order of the Cabibbo angle.

By keeping only the neutral Higgs components ($\Phi^0_i$) the effective Lagrangian is written as,
\begin{eqnarray}
- {\cal{L}}_Y &=&    Y^u_{ij}\bar{u}_i u_j  \Phi^0_1  + Y^d_{ij} \bar{d}_i d_j \Phi^0_2
                   + Y^l_{ij} \bar{l}_i l_j \Phi^0_3 + \nonumber \\
           &  &  [ Z^u_{ij} \bar{u}_i u_j \Phi^0_1  + Z^d_{ij} \bar{d}_i d_j \Phi^0_2 +
  Z^l_{ij} \bar{l}_i l_j \Phi^0_3 ] \frac{1}{u} (s_1+is_2)  + h.c.       ,          
\end{eqnarray}
where the Yukawa matrix is given as
\begin{eqnarray}
Y^f_{ij} = \rho^f_{ij} (\lambda_F)^{n^f_{ij}},
\end{eqnarray}
while the flavon interactions are described by the matrix
\begin{eqnarray}
	Z^f_{ij} = \rho^f_{ij} n^f_{ij} (\lambda_F)^{n^f_{ij}}.
\end{eqnarray} 

Substitution of 
\begin{eqnarray}
	 \Phi^0_i = \frac{1}{\sqrt{2}} (v_i + \phi^0_i + i \chi^0_i),
\end{eqnarray}
leads to identifying the mass terms and the Yukawa and flavon interactions. We need to diagonalize
the fermion mass matrices, by the usual bi-unitary tranformations. Here we assume the Yukawa matrices
are symmetric. Furthermore, we only write here 
the interaction Lagrangian for the real components of the neutral Higgs fields,
\begin{eqnarray}
 -{\cal{L}}_Y &=&  \bar{U} \frac{M_u}{\sqrt{2} v_1} U \phi^0_1 + 
                \bar{D} \frac{M_d}{\sqrt{2} v_2} D \phi^0_2
             +  \bar{L} \frac{M_l}{\sqrt{2} v_3} L \phi^0_3 \nonumber \\
            &  & +  [ \frac{v_1}{u}  \bar{U} \tilde{Z}^u U  + \frac{v_2}{u} \bar{D} \tilde{Z}^d D 
  + \frac{v_3}{u}  \bar{L}  \tilde{Z}^l L ] s_1   \nonumber \\
   &  &  + [  \bar{U} \tilde{Z}^u U \phi^0_1  + 
                    \bar{D} \tilde{Z}^d D \phi^0_2 
  + \bar{L}  \tilde{Z}^l L \phi^0_3 ] \frac{1}{u} s_1  + h.c. \; .
\end{eqnarray}
The matrices $\tilde{Z}^f$ are written now in the mass-eigenstate basis, and in general are
not diagonal, thus they will induce FCNC mediated by the SM-like Higgs field. 
The capital letters for (Dirac) fermion fields are used to indicate the vector of
mass eigenstates, i.e. $U_i=(u,c,t)$, $D_i=(d,s,b)$, and $L_i=(e,\mu,\tau)$.

\subsection{The scalar potential}

The scalar potential for the 2 + 1 model was studied in \cite{Grzadkowski:2010au},
where it was found that the potential of the model allows a minimum that satisfies the
charge-color-breaking and unbounded-from-below conditions. 
On the other hand, the potential for the inert dark matter model with a complex singlet 
was analised in \cite{Bonilla:2014xba}; this model is a kind of 1 + 1 + FN singlet model.
In our particular case, our model is a mixture of the above scenarios. The complete expression for the scalar potential can be found in the Appendix \ref{ScalarPotential}.

{
The study of the vacuum structure for multi-Higgs doublet models, which started with the work of \cite{DiazCruz:1992uw}, can be always analyzed from the results of the  
three Higgs doublet case with the proper choice of basis \cite{Barroso:2006pa}. It is a remarkable result which nevertheless
substantially departs from the two Higgs doublet scenario, namely: whenever there
exists a normal minimum at tree level it will always be below any Charge-Breaking (CB) stationary point \cite{Ferreira:2004yd,Barroso:2005sm}, leaving $U(1)_{em}$ invariant. This property is lost for three or more Higgses.
There is a sufficient
condition though by which one can guarantee that the global minimum leaves untouched the conservation of electric charge within an $N$-Higgs doublets scenario. Basically, this condition demands that the parameters of the potential are such that after reaching the so called B-basis the normal vacuum structure mimics that of a two Higgs doublet model \cite{Barroso:2005sm}. The introduction of a gauge singlet (flavon) field does not alter the previous picture. Thus, one can at least choose regions of parameters with the correct vacuum structure. 

Recall that} the scalar content of the model includes three ``active" Higgs doublets ($\Phi_i$) and one
inert doublet ($\Phi_4$), as well as, a singlet, $S$. {Now,} these fields are written as follows
\begin{eqnarray}
	\Phi_i = \begin{pmatrix}
	\varphi_i^+ \\
	\frac{v_i + \phi_i^0 + i \chi_i^0}{\sqrt{2}}
	\end{pmatrix}, \; (i=1,2,3) \qquad
	\Phi_4 = \begin{pmatrix}
	s^+ \\
	\frac{s^0 + i P^0}{\sqrt{2}}
	\end{pmatrix},
\end{eqnarray}
and
\begin{eqnarray}
	S = \frac{1}{\sqrt{2}}(u + s_1 + i s_2).
\end{eqnarray}

\begin{figure}[H]
\centering
\includegraphics[width=3.5 in]{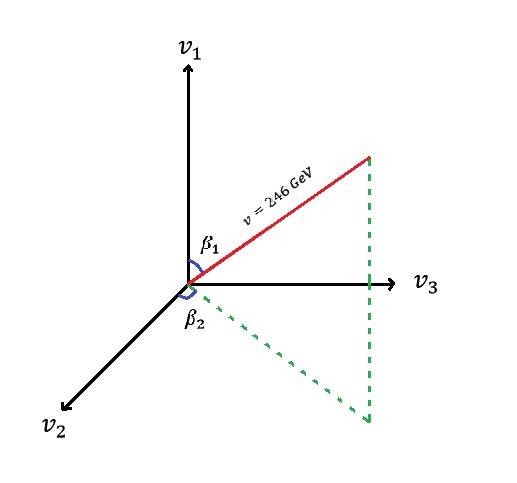}
\caption{The Higgs vevs in spherical coordinates.
\label{vevs-spherical}}
\end{figure}

The Higgs and flavon fields are written in terms of the mass eigenstates through the
rotation  $O^T$ of dimensions $(4\times 4)$: 
\begin{eqnarray}
Re(\Phi^0_i) &=&  O^T_{i1} h^0_1 + O^T_{i2} H^0_2 + O^T_{i3} H^0_3 + O^T_{i4} H^0_F, \nonumber \\
Re (S)       &=&  O^T_{41} h^0_1 + O^T_{42} H^0_2 + O^T_{43} H^0_3 + O^T_{44} H^0_F.
\end{eqnarray}
The details of the diagonalization depend on the Higgs potential, however as one can see from
the discussion of refs. \cite{Ivanov:2012fp,Keus:2013hya}, there are plenty of parameters
to have a reasonable particle spectrum, including the decoupling limit,
such as the one we shall discuss here.
Furthermore, as the vev's must satisfy: $v^2_1+ v^2_2+v^2_3 = v^2$, with $v=246$ GeV, 
we find convenient to use spherical coordinates to express each vev ($v_i$) in terms of the
total vev $v$ and the angles $\beta_1$ and $\beta_2$, as shown in Fig. \ref{vevs-spherical}, i.e.,
$v_1 = v \cos\beta_1$, $v_2 = v \sin \beta_1 \cos \beta_2$ and 
$v_3 = v \sin \beta_1 \sin \beta_2$. 

{There is an important issue which needs to be discussed here. It emerges from the use of the FN mechanism; particularly, from the nature of the employed flavor symmetry. Within a $U(1)$-type FN model, the Abelian flavor symmetry can be either local or global. When it is local, it includes a gauge boson, which becomes massive after the flavon develops a vev ($u$). This boson has both FC and FV interactions and its phenomenology severely restricts its mass, that turns out to be $\frac{1}{g_{Z'}} M_{Z'} \gtrsim 10-100 \text{ TeV}$ \cite{Tsumura:2009yf} and which would be out of the LHC reach. 
Then, we come to the scalar sector which could be seen as an effective approach \cite{Tsumura:2009yf}. On the other hand, in the minimal model with a global flavor symmetry and one flavon (complex singlet under the SM) there are two d.o.f., when CP is conserved in the Higgs sector the FN singlet includes one CP-even scalar and one CP-odd. Because of the SSB of the flavor symmetry, the CP-odd would be a massless NGB,  while the CP-even would be heavy. To have a viable model one breaks explicitly the global symmetry with a mass term ($M^2$) and the CP-odd becomes a PGB. Their masses could be written as,
	\begin{eqnarray}
		M_P^2 = M^2, \hspace{2 cm} M_{H_f}^2 = M^2 + \lambda' u^2.
	\end{eqnarray}
	As one of our goals is to study FV decays of the SM-like Higgs, we could use a mild amount of fine-tunning to have $M_P \approx M_{H_f}$.
	A more "natural" option to transmit LFV to the SM-like Higgs would be to consider CPV in the scalar sector, then it would be allowed to have a mixing of the SM-like Higgs and the CP-odd component of the FN singlet. In this case, the origin of LFV interactions would be different from the one considered in our paper. However, the coupling of the SM Higgs implications for the Higgs phenomenology would be qualitatively (and even quantitatively) similar. Thus, in the following, we shall focus in the phenomenological implications arising from the light SM-like Higgs boson.}

Although a detailed discussion of the scalar spectrum requires a study of the potential, we can
advance some general properties of the spectrum. Each of these states will show some peculiar manifestations from their interaction with
the flavor sector, which include:

\begin{itemize}

\item {\bf{A SM-like Higgs state with $m_h=125-126$ GeV:}}  The scalar sector must include
 a lightest state that would be identified with the SM-like Higgs boson. Here one expects, as
 the imprints of the model,  the appearance of
 small deviations from SM predictions for the  diagonal Higgs-fermion couplings. 
 Furthermore, it will be possible to induce new FV Higgs
 couplings. 
 To study these effects it could be enough to work within a model that contains one Higgs doublet 
 and one flavon field. However, in this case the  couplings of the Higgs boson with all fermion types 
 (up-, d-type quarks and leptons) as function of mass should lay on a single line, as in the SM.
 
\item {\bf{States with flavon-dominated composition:}} These states, which could be called
 flavons for short, could provide the more radical signature of the models under consideration.
 The observation of these signals depends on the flavon mass scale, and could be at the reach of
 the LHC if such scale was about $O(5-10)$ TeV. It would be interesting to learn what could be
 the reach of a 100 TeV collider \cite{Arkani-Hamed:2015vfh}. To study the most relevant flavon effects it could also be enough 
 to consider  a model that contains one Higgs doublet and one flavon field.

\item {\bf{Heavy Higgs bosons with large flavon mixing:}} In multi-Higgs models, the Higgs spectrum 
will also include extra heavy Higgs states that could have significant mixing with the flavon fields. 
The properties of these heavy Higgs bosons could deviate significantly from SM expectations and
can have interesting properties, such as large FV couplings, 
that could also be searched at the LHC. 
The behavior of couplings as function of the mass, will depend on the Higgs content and type of
model used. For instance within the so called 2HDM of type II, 
one Higgs doublet couples to up-type quarks, while the second Higgs doublet couples to down-type 
quarks and leptons. In this case the fermion couplings would lay on two lines:
one for up-type quarks and another one for d-type quarks and leptons. How distinguishable these lines
could be, will depend on how close the model is to the decoupling limit.
We would need to consider a model with three Higgs doublets, with  one 
Higgs doublet being associated for each fermion type, in order to have the fermion couplings 
with the lightest SM like Higgs boson lying on three different lines. 

\item {\bf{Inert Higgs states:}} On the other hand, with the inclusion of an extra Higgs doublet 
of the Inert type, one could also account for the Dark Matter of the universe.
A new feature of this case is the couplings of the Higgs bosons with the DM candidate,
which could manifest in the form of invisible decays. Although the detailed properties will depend
on the Higgs potential, the essential information will be contained in the mass of the DM candidate
and its coupling with the SM-like Higgs boson, which will allow to calculate the invisible
Higgs decay and compare with the LHC limits on the invisible Higgs branching ratio.

\end{itemize}

\begin{figure}[H]
\centering
\includegraphics[width=2.1 in]{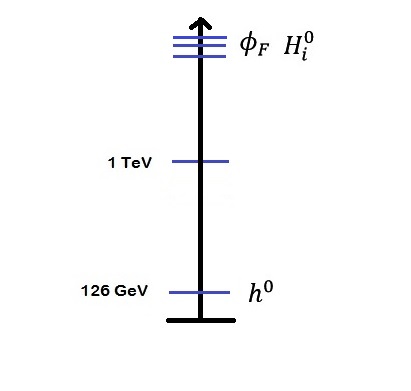}
\includegraphics[width=2.1 in]{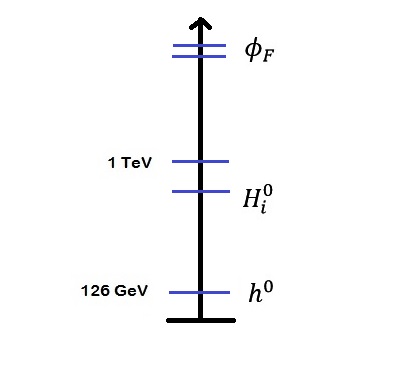}
\includegraphics[width=2.5 in]{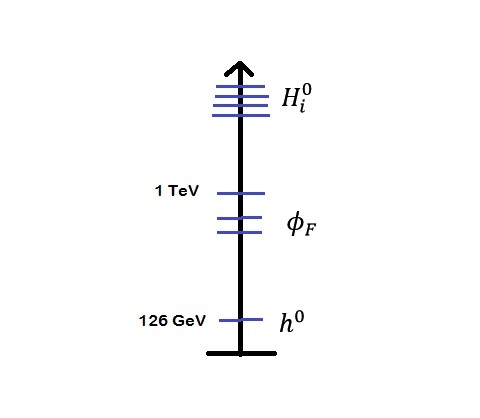}
\caption{The Higgs and flavon spectrum in multi-Higgs models.
\label{spectra-types}}
\end{figure}

These possibilities for the spectrum of Higgs bosons and flavons are illustrated in Fig. \ref{spectra-types}.
In the first case, we have only one SM-like Higgs boson at low energies, i.e. with $m_h \simeq 125-126$ GeV, 
with the heavy Higgses and flavons having similar masses, which could be of order O(TeV) up to the multi-TeV range. 
In the second case, only the heavy Higgs bosons have masses below the TeV range, with some detectable remnant effects 
being left from their mixing with the  flavon fields. Finally, a more exotic scenario would include  a light flavon field, 
with  mass below the O(TeV), which in itself could provide interesting discovery signals, but it
could also mix significantly with the SM-like Higgs boson, and induce large FV Higgs couplings.

Apart from the Yukawa couplings, we also need to specify the Higgs coupling with the vector bosons,
which is written as $g_{hVV} =  g^{sm}_{hVV} \chi_V$, with the factor $\chi_V$ 
given as,
\begin{eqnarray}
 \chi_V &=&  \frac{v_1}{v}O^T_{11} + \frac{v_2}{v}O^T_{21}+ \frac{v_3}{v}O^T_{31}, \nonumber \\
     &=&  \cos\beta_1 \, O^T_{11} + \sin\beta_1 \cos\beta_2 \, O^T_{21}+ \sin\beta_1 \sin\beta_2 \, O^T_{31}.
\end{eqnarray}

It is interesting to notice that the coupling $\chi_V$ can be  written in terms of the  fermionic couplings,
which can be seen as a type of sum rule, i.e.,
\begin{equation}
 \chi_V = \cos^2\beta_1 \, \eta^u + \sin^2\beta_1 \cos^2\beta_2 \, \eta^d + \sin^2\beta_1 \sin^2\beta_2 \, \eta^l.
\end{equation}

\subsection{The structure of the $Y$ and $Z$  matrices and the Higgs couplings}

The interaction Lagrangian for the Higgs-fermion couplings can be expressed in
terms of the lightest Higgs state, which is identified as $h^0=h^0_1$, 
\begin{equation} \label{Eff-Higgs-Yukawa}
 -{\cal{L}}_Y =  [ \frac{\eta^u } {v} \bar{U} M_u U  + \frac{\eta^d  }{v} \bar{D} M_d  D  
             + \frac{\eta^l }{v}  \bar{L} M_l L  
            +  \kappa^u \bar{U}_i \tilde{Z}^u U_j  + \kappa^d \bar{D}_i \tilde{Z}^dD_j 
                + \kappa^l   \bar{L}_i\tilde{Z}^l  L_j ] h^0  + h.c.\; ,
\end{equation}
where $\eta^u = O^T_{11}/\cos\beta_1$, $\eta^d=O^T_{21}/\sin \beta_1 \cos \beta_2 $, and
$\eta^l=O^T_{31}/ \sin \beta_1 \sin \beta_2$, describe the strength of the flavor-diagonal
Higgs couplings. While the FV Higgs couplings are described by the parameters:
$\kappa^u = \frac{v}{u} O^T_{41} \cos\beta_1 $, \, 
$\kappa^d = \frac{v}{u} O^T_{41} \sin\beta_1 \cos \beta_2 $, \, and
$\kappa^l = \frac{v}{u} O^T_{41} \cos\beta_1 \sin \beta_2 $.

We do not discuss in detail the charge assignments under the Abelian flavor symmetry
as many choices have appeared in the literature, e.g. Hermitian, non-Hermitian, etc, {see for example \cite{Dery:2014kxa,Huitu:2016pwk,Dreiner:2003yr}}.  
Rather, we will present one form of a viable Yukawa matrix to be explored here and consider the case where there is only one scale $\Lambda_F$; implying
$\lambda \equiv \lambda_F = 0.22$. Namely, for the up-type quarks we assume
\begin{equation}
Y^{u}=
 \left( \begin{array}{ccc}
 \rho^u_{11} \lambda^{4}  &   \rho^u_{12} \lambda^{4}     &    \rho^u_{13} \lambda^{4}  \\
\rho^u_{21} \lambda^{4}   &   \rho^u_{22} \lambda^{2}     &    \rho^u_{23} \lambda^{2}  \\
\rho^u_{13} \lambda^{4}    &   \rho^u_{23} \lambda^{2}      &    \rho^u_{33}   
\end{array} \right).
\end{equation}
Very importantly, due to
 the form of the Higgs/flavon couplings, the 33 entry does not have
a power of $\lambda$, which means that the flavon coupling with the top quark will be suppressed
and will be of the same order as the coupling to charm quarks or the FV Higgs coupling with $tc$.

On the other hand, we consider the Yukawa matrix of both the charged leptons ($Y^l$) and d-type quarks
to have a similar structure,
\begin{equation}
Y^{d}=
 \left( \begin{array}{ccc}
 \rho^d_{11} \lambda^{6}    &   \rho^d_{12} \lambda^{6}     &    \rho^d_{13} \lambda^{6}  \\
 \rho^d_{21} \lambda^{6}    &   \rho^d_{22} \lambda^{4}     &    \rho^d_{23} \lambda^{4}  \\
 \rho^d_{13} \lambda^{6}    &   \rho^d_{23} \lambda^{4}     &    \rho^d_{33} \lambda^{2}  
\end{array} \right).
\end{equation}
For leptons, we change $\rho^d_{ij} \to \rho^l_{ij}$. The choice for powers of $\lambda$ is not unique,
and in fact it could also change when the  vevs ($v_i$) of the Higgs doublets have a hierarchical
structure. From this points onwards, we only focus in the implications arising from flavon-Higgs mixing effects in the quark sector and leave aside their study in the charged lepton sector.

Now, notice the following simplification. As a consequence of the hierarchical structure in the mass matrices we can study
two particular cases: first, the subsystem $2-3$
made out of the two heaviest families and second, the subsystem $1-2$ which emerges after diagonalization of the previous case and after taking the limit $m_3 \rightarrow \infty$. The former case is basically considering a rank two matrix with the first family being massless, as allowed by the Schmidt-Mirsky approximation
theorem, while the latter one is the decoupling limit where the mixing between the first
and third family has been neglected. 

\subsubsection{The rank two case or first family massless limit}
By construction, the couplings to the first
generation quarks and leptons are the most suppressed.
And this suppression, in return, allows us to use the rank two limit for the Yukawa matrices. For example, in the quark sector they take the form
\begin{eqnarray}
	Y^u \simeq  \left( \begin{array}{ccc}
 0  &   0     &    0 \\
0  &   \rho^u_{22} \lambda^{2}     &    \rho^u_{23} \lambda^{2}  \\
0   &   \rho^u_{23} \lambda^{2}      &    \rho^u_{33}  
\end{array} \right),
\end{eqnarray}
and
\begin{equation}
Y^{d}\simeq
 \left( \begin{array}{ccc}
0    &   0     &    0 \\
 0    &   \rho^d_{22} \lambda^{4}     &    \rho^d_{23} \lambda^{4}  \\
 0    &   \rho^d_{23} \lambda^{4}     &    \rho^d_{33} \lambda^{2}  
\end{array} \right).
\end{equation}

Now, in order to study the FV Higgs couplings we need to consider the 2-3 sub-system given by the matrices $\tilde{Z}^f$. For the up- and down-type quarks the corresponding $Z^f$-submatrices, written in
the quark mass eigenstate basis,  are
\begin{equation}
\tilde{Z}^{u} \simeq
 \left( \begin{array}{cc}
 2 Y^u_{22}      &    2 Y^u_{23}   \\
 2 Y^u_{23}      &    4 s_u Y^u_{23}   
\end{array} \right),
\end{equation}
and
\begin{equation}
\tilde{Z}^{d} \simeq
 \left( \begin{array}{cc}
  4 Y^d_{22}    &    2 Y^d_{23}  \\
  2 Y^d_{23}   &   2 Y^d_{33}   
\end{array} \right).
\end{equation}
The above matrices can be found by first performing a diagonalization of the Yukawa matrices to leading order in the individual mixing angles ($s_{u,d}$) and then, by virtue of these now known rotations, transform the $Z^f$ matrices. Through this procedure we are able to find a relation among
the parameters such that we can express the $\rho^{u,d}_{ij}$'s in terms of the quark mass ratios
and the CKM angle $|V_{cb}| \simeq s_{23}$. 

Specifically, let us define: 
$r_u= m_c/m_t$, 
$r^u_1= Y^u_{22}/Y^u_{33}$, and $r^u_2= Y^u_{23}/Y^u_{33}$. 
Similarly, $r_d= m_s/m_b$,
$r^d_1= Y^d_{22}/Y^d_{33}$ and $r^d_2= Y^d_{23}/Y^d_{33}$. 
Now, the diagonalization procedure involves the approximated eigenvalues
\begin{eqnarray}
	Y_{33}^f \simeq \frac{m_{f,3}}{\omega_f} \hspace{.5 cm} \text{and} \hspace{0.5 cm} 
	Y_{22}^f - 2 s_f Y_{23}^f + s_f^2Y_{33}^f \simeq \frac{m_{f,2}}{\omega_f} ,
\end{eqnarray}
with $\omega_u = v_1/\sqrt{2}$, $\omega_d = v_2/\sqrt{2}$, and $f=u,\;d$. Recall that the mixing matrix is being built by contributions coming only from the $2-3$ sector, $V_\text{CKM} \simeq L_{23}^u (L_{23}^d)^\dagger$, where $L_{23}^f$ is a unitary matrix transforming the corresponding left handed field and only acting in the 2-3 family subspace. It is then straightforward to show that the relation between the mixing angle and the individual ones,
\begin{eqnarray}
	s_{23} \simeq s_u - s_d,
\end{eqnarray}
allows one to find
\begin{equation}
r_1^u \approx r_1^d + r_u - r_d + 2s_{23} \sqrt{r_1^d - r_d} + s_{23}^2.
\end{equation}
Thus, we can either vary $r^d_1$ and get values for $r^d_2 \equiv \sqrt{r_1^d-r_d}$ or vice-versa, and then from the above expression find $r^u_1$ and with it $r^u_2 \equiv \sqrt{r_1^u-r_u}$. Also, to get 
the $\tilde{Z}$ elements, we need to express the Yukawas $Y_{33}$ in terms of the quarks masses,
that is:
$Y^u_{33} = \sqrt{2} m_t/ v_1$ and  $Y^d_{33} = \sqrt{2} m_b/ v_2$.
Finally, we also need to specify the values of $v_1$ and $v_2$; a few interesting scenarios 
are defined in the next sub-section.

\subsubsection{The third family decoupling limit}
After diagonalization of the 2-3 subsystem, we can neglect mixing between the first
and third generation. In return, we may only focus on the 1-2 subsystem. The corresponding
$Z$-submatrices, written in the quark mass eigenstate basis, are
\begin{eqnarray}
	\widetilde{z}^u \simeq
 \left( \begin{array}{cc}
  4 Y^u_{11}    &    2 Y^u_{12}  \\
  2 Y^u_{12}   &   2 Y^u_{22}   
\end{array} \right),
\end{eqnarray}
and
\begin{eqnarray}
	\widetilde{z}^d \simeq
 \left( \begin{array}{cc}
  4 Y^d_{11}    &    2 Y^d_{12}  \\
  2 Y^d_{12}   &   2 Y^d_{22}   
\end{array} \right).
\end{eqnarray}

Finally, an expression relating the Cabibbo angle with these Yukawa couplings
can be obtained as in the previous case
\begin{equation}
p_1^u \approx p_1^d + p_u - p_d + 2s_{12} \sqrt{p_1^d - p_d} + s_{12}^2,
\end{equation}
where $p$ means exchanging the $2$ and $3$ indices in the $r$ parameters by the corresponding $1$ and $2$.

\subsection{Parameter scenarios}

In order to study the predictions of our model, we need to specify the vevs $v_i$ and the 
rotation matrix for Higgs particles ($O_{ij}$). We can take inspiration from the 2HDM results
(see for instance \cite{Ferreira:2014naa}), which show that the LHC Higgs data favors both the
decoupling and  alignment solutions, that is to say, both $\tan\beta >> 1$ and $\tan\beta \simeq 1$  
are acceptable solutions. In terms of the Higgs vevs, this means that they are either of the same 
order or one is much larger than the other.
Thus, for the vev's we use the spherical coordinates ($\beta_1$ and $\beta_2$), and  
leave $\beta_1$ as a free parameter and then explore the following cases:
\begin{itemize}
 \item (VEV1) We can take  first $v_2=v_3$, which in  spherical coordinates, means:
        $\beta_2=\frac{\pi}{4}$, and any value of $\theta$,
 \item (VEV2) We also consider unequal vevs with $v_2<v_3$, which means $\beta_2=\frac{\pi}{3}$,
 \item (VEV2) We also consider unequal vevs with $v_2>v_3$, which means $\beta_2=\frac{\pi}{6}$,
 \end{itemize}

 Then, for the rotation matrix $O$ of real components of scalar fields, we can identify several 
 interesting scenarios:
 
 \begin{enumerate}
  \item The case where the 126-Higgs is   lighter than the heavy Higgs particles and
        the flavons,  i.e.  $m_h <  m_{H_i} \simeq m_{H_F}$, which have masses of order TeV.
   \item The case where the 126-Higgs is  lighter than the heavy Higgs particles, which in turn are 
         much lighter  than the flavons,
         i.e.  $m_h < m_{H_i} << m_{H_F}$,
  \item The case where the 126-Higgs is lighter than the flavons, which in turn are much lighter than the 
  heavy Higgs particles,
         i.e.  $m_h << m_{H_F} << m_{H_i}$,      
 \end{enumerate}
 
 Here, in order to relate the parameters, we take into account a special sub-case within the first case, which means assuming that 
 $O^T_{11} > O^T_{i1}$ and using the orthogonality relation for 
 the rotation matrix $O$
 \begin{equation}
  (O^T_{11})^2 +  (O^T_{21})^2+ (O^T_{31})^2+ (O^T_{41})^2 = 1.
 \end{equation}
 
Therewith, consideration of $O^T_{i1}\simeq O^T_{j1}$ (for $i\neq j$) leads one to have
 \begin{equation}
   O^T_{j1} = \sqrt{ \frac{1 -  (O^T_{11})^2}{3}}.
 \end{equation}

\section{Flavor conserving Higgs couplings at the LHC}
\label{Sec:IV}

The current data on Higgs production has been used to derive bounds on the Higgs couplings,
which describe the allowed deviation from the SM. In particular, ref. \cite{Giardino:2013bma} has
derived bounds on the parameters $\epsilon_X$, which are defined as the (small) deviations of the 
Higgs couplings from the SM values, i.e. $g_{hXX}= g^{sm}_{hXX} (1 + \epsilon_X)$.
We find very convenient, in order to use these results and
get a quick estimate of the bounds, to write our parameters as: $\eta^X= 1 + \epsilon_X$. 
For fermions, the allowed values are:
$\epsilon_t= -0.21\pm 0.22$, $\epsilon_b= -0.19\pm 0.30$, $\epsilon_{\tau}= 0.00 \pm 0.18$;
for the W and Z bosons these numbers are: $\epsilon_W= -0.15\pm 0.14$ and $\epsilon_Z= -0.01\pm 0.13$.

\begin{figure}[H]
\centering
\includegraphics[width=4 in]{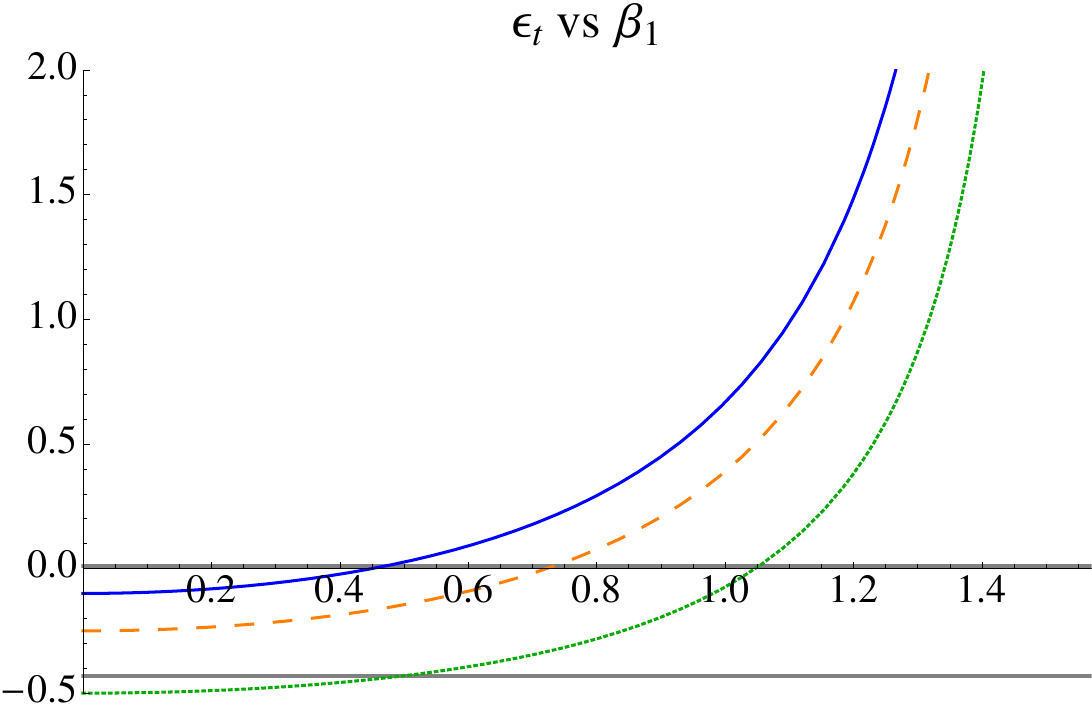}
\caption{
The deviations of the Higgs couplings from the SM values are defined by $g_{hXX}= g^{sm}_{hXX} (1 + \epsilon_X)$.  Here it is shown the small deviation, $\epsilon_t$, to the top quark as function of the angle $\beta_1$. The horizontal lines are the experimental limits on this factor, $\epsilon_t= -0.21\pm 0.22$ \cite{Giardino:2013bma}. The continous (blue) line,
large dashing (orange) line, and small dashing (green) line correspond to the cases
$\beta_2 = \pi/6$ and $O_{11} = 0.9$,  $\beta_2 = \pi/3$ and $O_{11} = 0.75$, and
$\beta_2 = \pi/4$ and $O_{11} = 0.5$, respectively.
\label{l2j}}
\end{figure}

\begin{figure}[H]
\centering
\includegraphics[width=4 in]{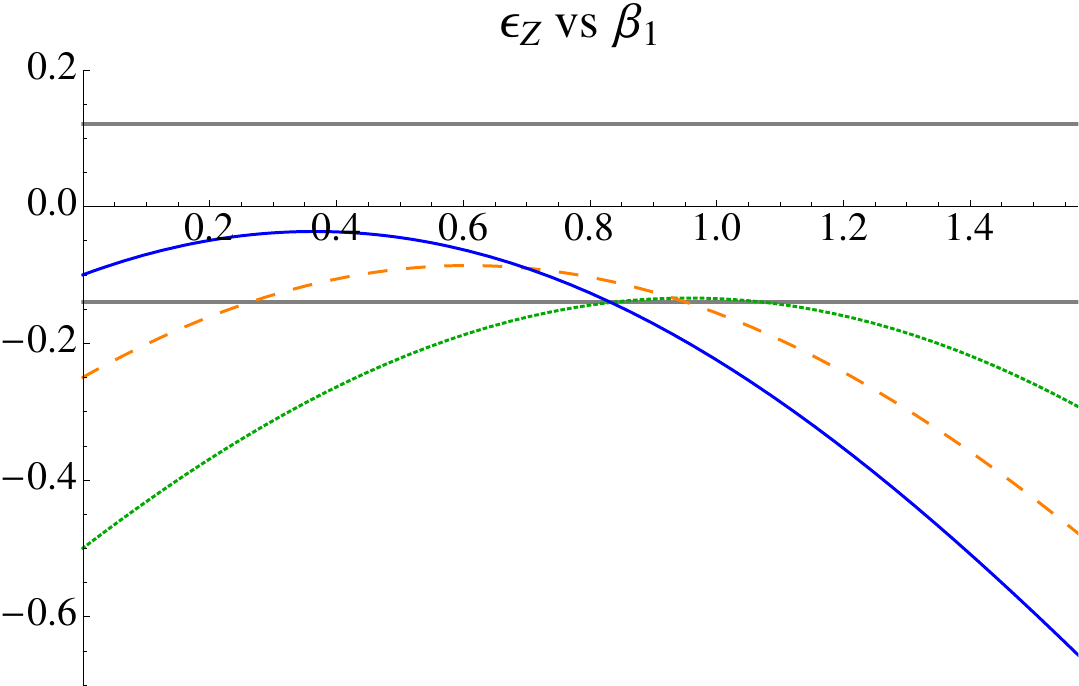}
\caption{
The deviations of the Higgs couplings from the SM values are defined by $g_{hXX}= g^{sm}_{hXX} (1 + \epsilon_X)$.  Here it is shown the small deviation, $\epsilon_Z$, to the Z boson as function of the angle $\beta_1$. The horizontal lines are the experimental limits on this factor, $\epsilon_Z= -0.01\pm 0.13$ \cite{Giardino:2013bma}. The continous (blue) line,
large dashing (orange) line, and small dashing (green) line correspond to the cases
$\beta_2 = \pi/6$ and $O_{11} = 0.9$,  $\beta_2 = \pi/3$ and $O_{11} = 0.75$, and
$\beta_2 = \pi/4$ and $O_{11} = 0.5$, respectively.
\label{l3j}}
\end{figure}

\begin{figure}[H]
\centering
\includegraphics[width=4 in]{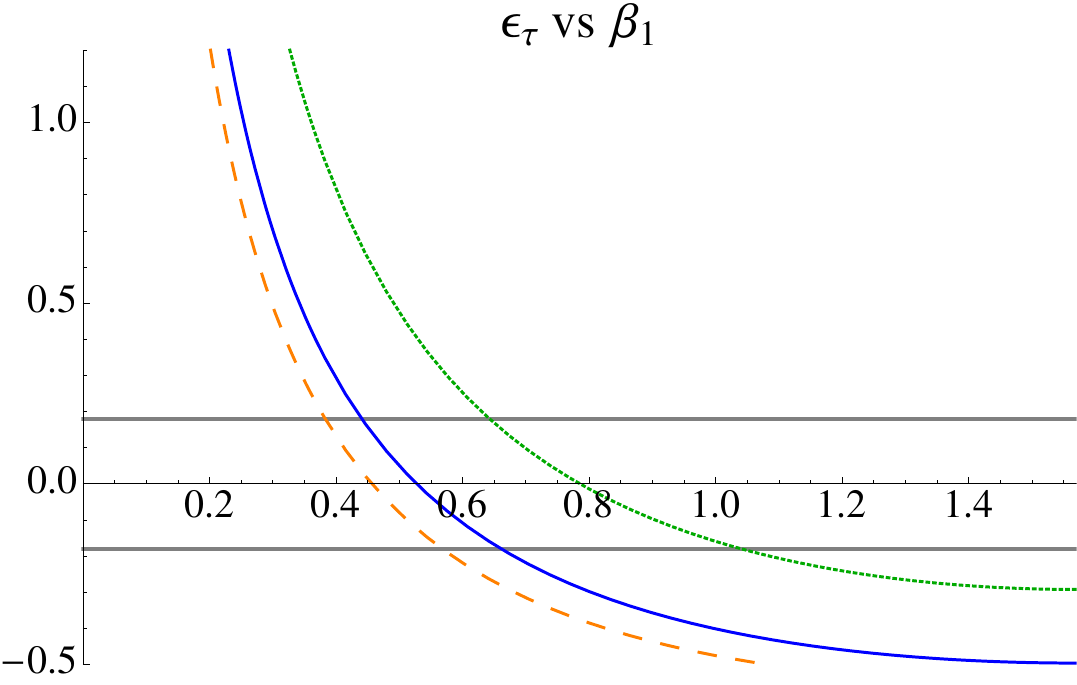}
\caption{
The deviations of the Higgs couplings from the SM values are defined by $g_{hXX}= g^{sm}_{hXX} (1 + \epsilon_X)$.  Here it is shown the small deviation, $\epsilon_\tau$, to the tau lepton as function of the angle $\beta_1$. The horizontal lines are the experimental limits on this factor, $\epsilon_{\tau}= 0.00 \pm 0.18$ \cite{Giardino:2013bma}. The continous (blue) line,
large dashing (orange) line, and small dashing (green) line correspond to the cases
$\beta_2 = \pi/6$ and $O_{11} = 0.9$,  $\beta_2 = \pi/3$ and $O_{11} = 0.75$, and
$\beta_2 = \pi/4$ and $O_{11} = 0.5$, respectively.
\label{l1j}}
\end{figure}

\begin{figure}[H]
\centering
\includegraphics[width=4 in]{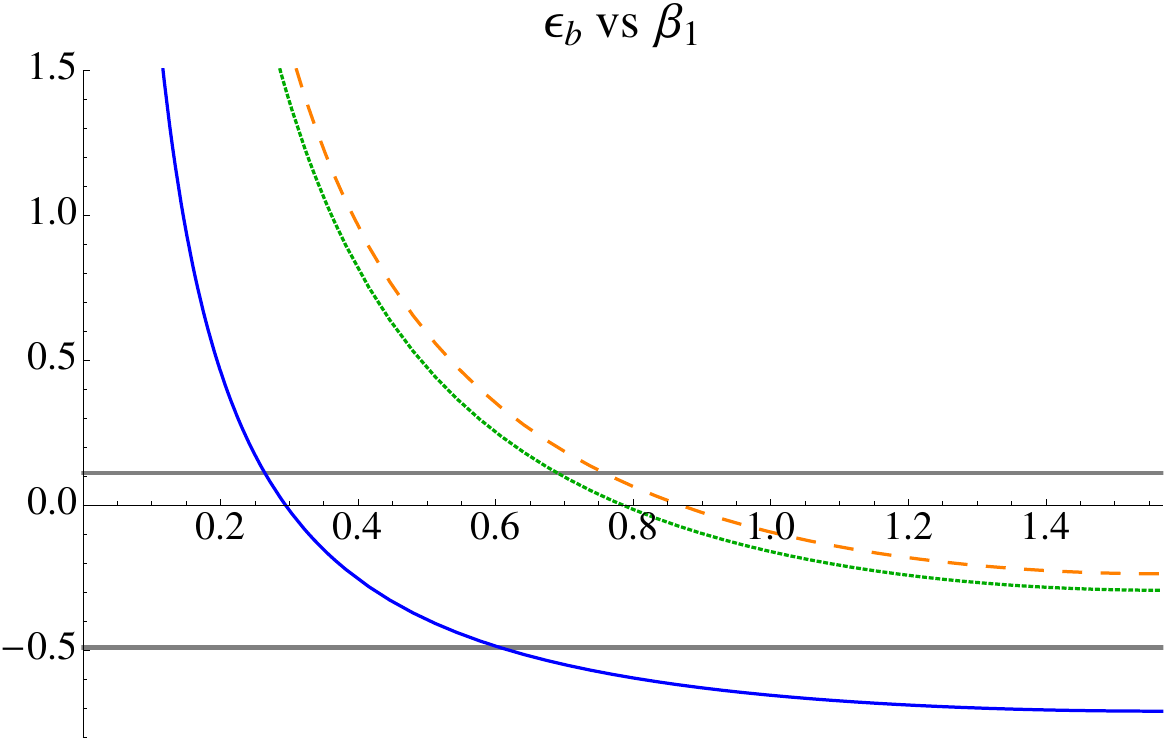}
\caption{
The deviations of the Higgs couplings from the SM values are defined by $g_{hXX}= g^{sm}_{hXX} (1 + \epsilon_X)$.  Here it is shown the small deviation, $\epsilon_b$, to the bottom quark as function of the angle $\beta_1$. The horizontal lines are the experimental limits on this factor, $\epsilon_b= -0.19\pm 0.30$ \cite{Giardino:2013bma}. The continous (blue) line,
large dashing (orange) line, and small dashing (green) line correspond to the cases
$\beta_2 = \pi/6$ and $O_{11} = 0.9$,  $\beta_2 = \pi/3$ and $O_{11} = 0.75$, and
$\beta_2 = \pi/4$ and $O_{11} = 0.5$, respectively.
\label{l4j}}
\end{figure}

An extensive analysis of parameters satisfying these bounds will be presented elsewhere, 
with detailed numerical scans; here we shall only pick a few specific points in parameter space, 
which satisfy the LHC bounds, and will help us to  understand qualitatively the behaviour 
of the model. These points will also be used in the next section in our analysis of FCNC top decays. 
Thus, we show in Figs. \ref{l2j}-\ref{l4j} the predictions for each of these parameters, as function of the 
angle $\beta_1$, for the case with  $\beta_2= \frac{\pi}{3}, \, \frac{\pi}{4}, \,\frac{\pi}{6}$, 
and for $O_{11} = 0.5, \, 0.75, \, 0.9$. We can see that it is possible to satisfy these bounds for all the
$\epsilon$'s. 


One specific point, in agreement with all data, is :  $\beta_1=0.5$ with $O_{11} =0.9$ and $\beta_2=\frac{\pi}{6}$. 
For these values we have:
$\eta^u=1.03$, $\eta^d=0.61$, $\eta^l=1.04$, and $\chi_V=0.96$. 
This shows that $h$ behaves very much SM-like, except for the coupling with d-type quarks. 
Interestingly, recent measurements of the Higgs coupling to the different SM particles \cite{Aad:2015gba} report a similar behaviour wherein the bottom quark shows a smaller coupling ($\eta^d_{\text{exp}}=0.63^{+0.39}_{-0.37}$) when compared to the rest which are larger than one.
Also, an important insight into the scalar structure of the SM, which will help us to discriminate between a multi-Higgs theory from a single one, will be gained when the Higgs couplings to the light-quarks have been measured at the LHC. These measurements are rather difficult and are requiring new techniques, see for example \cite{Bishara:2016jga, Soreq:2016rae}.

\begin{figure}[H]
\centering
\includegraphics[width=4 in]{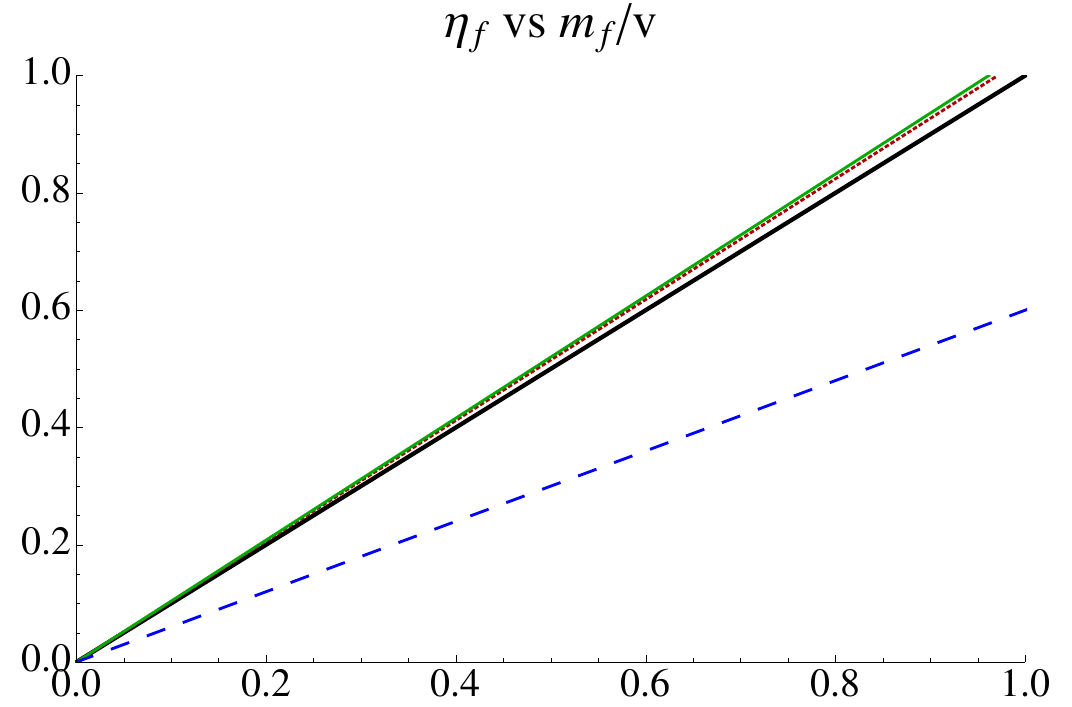}
\caption{ The Higgs-fermion coupling factors, $\eta_f$, as function of the normalized mass, $m_f/v$. 
The specific point $\beta_1=0.5$ with $O_{11} =0.9$ and $\beta_2=\frac{\pi}{6}$ is in
agreement to all data. For these values we have:
$\eta^u=1.03$, $\eta^d=0.61$, and $\eta^l=1.04$.
The bold continous (black) line, small dashing (red) line, large dashing (blue) line, and continous (green) line, correspond to the SM case and for the new set of couplings to the Higgs of the up-type quarks, d-type quarks, and charged leptons, respectively.
\label{fig:Hffcouplings}}
\end{figure}

Now, using the above mentioned set of values we can plot the Higgs-fermion coupling as function of the mass,
which is shown in Fig. \ref{fig:Hffcouplings}. 
We can see that the couplings for each fermion type lay on 
different lines, which could be distinguished from the SM (black line), and all of them 
are in agreement with the LHC Higgs data. Future measurements of these couplings at the ILC or FCC will help us in 
order to discriminate betwen our model predictions and those of the SM.

As it will be shown in the next section, the diagonal corrections contained
in the factors $\kappa^f \tilde{Z}^f$, will not change significantly the
above discussion for the top quark-Higgs couplings. However,
the Higgs coupling with the fermions ($b\bar{b}, c\bar{c}, \tau^+\tau^-$),
could be measured at the next-linear collider with a precision 
of a few percent, where it will be possible to test these effects. 
The corrections to the coupling $h\bar{b}b$, could
modify the dominant decay of the light Higgs,
as well as the associated production of the Higgs with 
b-quark pairs \cite{Balazs:1998nt, DiazCruz:1998qc}.

\section{FV Higgs couplings  and FCNC Top decays }
\label{Sec:V}

\subsection{Lessons from ${\cal{L}}_Y$ for FV Higgs couplings}

The Yukawa sector predicts the presence of non-diagonal couplings (in flavor space),
which can generate FCNC. However, we notice the following trends:
\begin{itemize}
 \item The appearance of the factor $v/u < 1$, brings a suppression of the FV couplings,
       as compared with the FC ones,
 \item In the limit $\cos\beta_1\to 1$, which is one option to get a light SM-like Higgs boson,
      the factor $\sin\beta_1 \to 0 $, which appears in the FV Higgs couplings, will also give 
       a suppression effect for d-type quarks and leptons, as compared with the top quark, which is 
       good because it will make easier to satisfy stronger bounds coming from $K-\bar{K}$ mixing and
       lepton FV transitions \cite{Buras:2010mh,Crivellin:2013wna},
 \item In general, the mixing-angle factors ($\eta$'s) that enhance (suppresses) the FC Higgs interactions,
       will produce a supression (enhancement) of the FV couplings ($\kappa$'s),
 \item The factor $O^T_{41} < 1$ will also induce another suppression effect,
 \item As we will show next, the Yukawa structure, which determines the form of the $\tilde{Z}$
       matrices, will also induce additional suppression effects.
\end{itemize}

\subsection{Constraints from $K$-$\bar{K}$ and $D$-$\bar{D}$ mixing}
Within our model we have flavor changing neutral current processes at tree level being effectively mediated by a SM-like Higgs boson, see Eq. \ref{Eff-Higgs-Yukawa}.
In order to avoid their dangerous possible large contributions to already measured observables
we apply strong contrains coming from
the study of neutral meson systems \cite{Nierste:2009wg}. In particular, the combination of K and D mixing gives
unavoidable bounds on new physics parameters \cite{Blum:2009sk}. 
The decoupling limit previously studied helps us to make these estimates. 

In general, the neutral K mass splitting is approximately given as
\begin{eqnarray}
	\Delta m_K \approx 2 |M_{12}|,
\end{eqnarray}
where $|M_{12}|$ is a measure of the small effects breaking the mass degeneracy in the
kaon system; the weak interaction, for example, is one of them. Experimentally, we have \cite{Agashe:2014kda}
\begin{eqnarray}
	\Delta m_K^{\text{exp}} = (3.484 \pm 0.006 )\; \mu\text{eV}.
\end{eqnarray}

On the other hand, using the vacuum insertion approximation one is able to obtain \cite{Branco:1999fs}
\begin{eqnarray}
	|M_{12}| = \frac{5 f_k^2 m^3_k (\kappa^d \widetilde{z}_{12})^2}{48(m_d +m_s)^2 m_H^2}.
\end{eqnarray}
From which we get the following upper bound
\begin{eqnarray}
	\kappa^d \widetilde{z}_{12}^d <  1 \times 10^{-6}.
\end{eqnarray}

Similarly, the experimental value for the mass difference in the D meson system is \cite{Barberio:2008fa}
\begin{eqnarray}
	\frac{\Delta m _D^{\text{exp}}}{m_D} = (8.6 \pm 2.1) \times 10^{-15},
\end{eqnarray}
from which we get the following upper bound 
\begin{eqnarray}
	\kappa^u \widetilde{z}^u_{12} <  1 \times 10^{-8}.
\end{eqnarray}

Even though these bounds are rather strong compared to the ones in the $2-3$ sector
they can actually be expected to be like that as by construction we already put some hierarchy among them. 

\subsection{Numerical choice for the $\tilde{Z}$ parameters}

Let us consider the following sample values: $r_2^d= 0.05, 0.1,$ and $0.3$, and also
assume: $\beta_1 = 0.5$, then Table \ref{Zmatrix} shows the values
of the entries for the $\tilde{Z}^u$ matrix for the second and third family case. 
We choose to focus 
on the up-quark sector, because we want to get an estimate for the most relevant
predictions of the model, which we believe is related to the top quark physics,
and in particular to the decay $t\to ch$.

For the specific point in parameter space presented in the previous section:
 $\beta_1=0.5$ with $O_{11} =0.9$ and $\beta_2={\pi}/{6}$, 
which is in agreement with the LHC data,
we obtain the following value $\kappa^u=0.22 \frac{v}{u}$. And with the entries
$\tilde{Z}^u$ shown in Table \ref{Zmatrix}, we obtain that the correction to the coupling 
$h\bar{c}c$ could be of order 50 percent for $\frac{v}{u}=0.1$, which could give some 
enhancement for the Higgs production through charm fusion; this could be worth further 
studying. On the other hand, the correction to  the top quark-Higgs coupling is
less than 0.1 percent, and thus negligible. 

\begin{center}
\begin{table}
\begin{center}
\begin{tabular}{| c | c | c | c | c |}
\hline
 Scenario  & $r^d_2$ & $\tilde{Z}^u_{22}$     &  $\tilde{Z}_{23}^u$ & $\tilde{Z}^u_{33}$  \\ 
 \hline \hline
  X      & 0.05  & $2.6 \times 10^{-2}$  & $2.0 \times 10^{-1}$  & $3.7 \times 10^{-2}$  \\ 
 \hline
  Y      & 0.1   & $5.2 \times 10^{-2}$  & $3.2 \times 10^{-1}$  & $8.8 \times 10^{-2}$  \\ 
  \hline
  Z      & 0.3   & $2.6 \times 10^{-1}$  & $7.6 \times 10^{-1}$  & $5.2 \times 10^{-1}$   \\
\hline
\end{tabular}
\end{center}
\caption{The entries 22, 23, and 33 for the matrix $\tilde{Z}^u$, after taking  the particular case of $\beta_1 = 0.5$.}
\label{Zmatrix}
\end{table}
\end{center}

\subsection{The FCNC decay $t\to ch$ }

Top quark rare decays will enter into a golden era, because the LHC is a top factory, with 
about $6 \times 10^6$ top pairs produced so far at the LHC. Thus, it is very
interesting to look at the top decays that have some potential to be detectable. The first 
calculation of FCNC top decays at one-loop, was for $t\to c\gamma$  \cite{LDCtopfv}, while the complete
calculations of the FCNC modes $t \to cX$ ($X=\gamma,g,Z,h$) was reported in ref. \cite{Eilam:1991}.
Then, the calculation for the mode $t \to ch$, was corrected in \cite{Mele:1998,Aguilar:2003}. 
In turn, the FCNC decays into pairs of vector bosons $t \to cW^+W^- (ZZ,\gamma\gamma)$, was presented in
\cite{Jenkins:1996zd,Bar-Shalom:1998,Bar-Shalom:2005,DiazCruz:1999ab}. The mode 
$t \to c\ell^-\ell^+$ was discussed in \cite{Frank:2006, DiazCruz:2012xa}. 
All of these modes have an extremely suppressed Branching Ratio (B.R.) within the SM, but they have it enhanced 
when New Physics  (NP) is involved,  see for instance \cite{Larios:2006}. 
More promising, in terms of having a larger B.R. the mode $t\to bW\ell^-\ell^+ $, which was studied recently \cite{Quintero:2014lqa}, fulfills this task.

From now on, we focus on the mode $t\to ch$, which can reach large B.R.'s. The decay width is given by
\begin{equation}
 \Gamma (t\to ch) = \frac{m_t}{6 \pi}  |\kappa^u \tilde{Z}^u_{23}|^2.
\end{equation}

\begin{center}
\begin{table}
\begin{center}
\begin{tabular}{| c| c | c | c |}
\hline
 Scenario  & u[TeV] & $\kappa^u \times \tilde{Z}^u_{23}$ & $B.R.(t\to ch)$  \\ 
 \hline \hline
   X1     & 0.5  & $2.1 \times 10^{-2}$    & $8.6 \times 10^{-5}$  \\ 
 \hline
   X2     & 1    & $1.1 \times 10^{-2}$    & $2.2 \times 10^{-5}$  \\ 
 \hline
   X3     & 10   & $1.1 \times 10^{-3}$    & $2.2 \times 10^{-7}$  \\ 
 \hline
   Y1     & 0.5  & $3.5 \times 10^{-2}$    & $3.5 \times 10^{-4}$  \\ 
 \hline
   Y2     & 1    & $1.7 \times 10^{-2}$    & $8.6 \times 10^{-5}$  \\ 
 \hline
   Y3      & 10  & $1.7 \times 10^{-3}$    & $8.6 \times 10^{-7}$  \\ 
 \hline
   Z1     & 0.5  & $8.3 \times 10^{-2}$    & $3.0 \times 10^{-3}$  \\ 
 \hline
   Z2     & 1    & $4.2 \times 10^{-2}$    & $7.8 \times 10^{-4}$  \\ 
 \hline
   Z3      & 10  & $4.2 \times 10^{-3}$    & $7.8 \times 10^{-6}$  \\ 
 \hline  
\end{tabular}
\end{center}
\caption{The factor $\kappa^u \times \tilde{Z}^u_{23}$ and branching ratios (B.R.) for $t\to ch$.}
\label{Tchbr}
\end{table}
\end{center}

\begin{figure}[H]
\centering
\includegraphics[width=4 in]{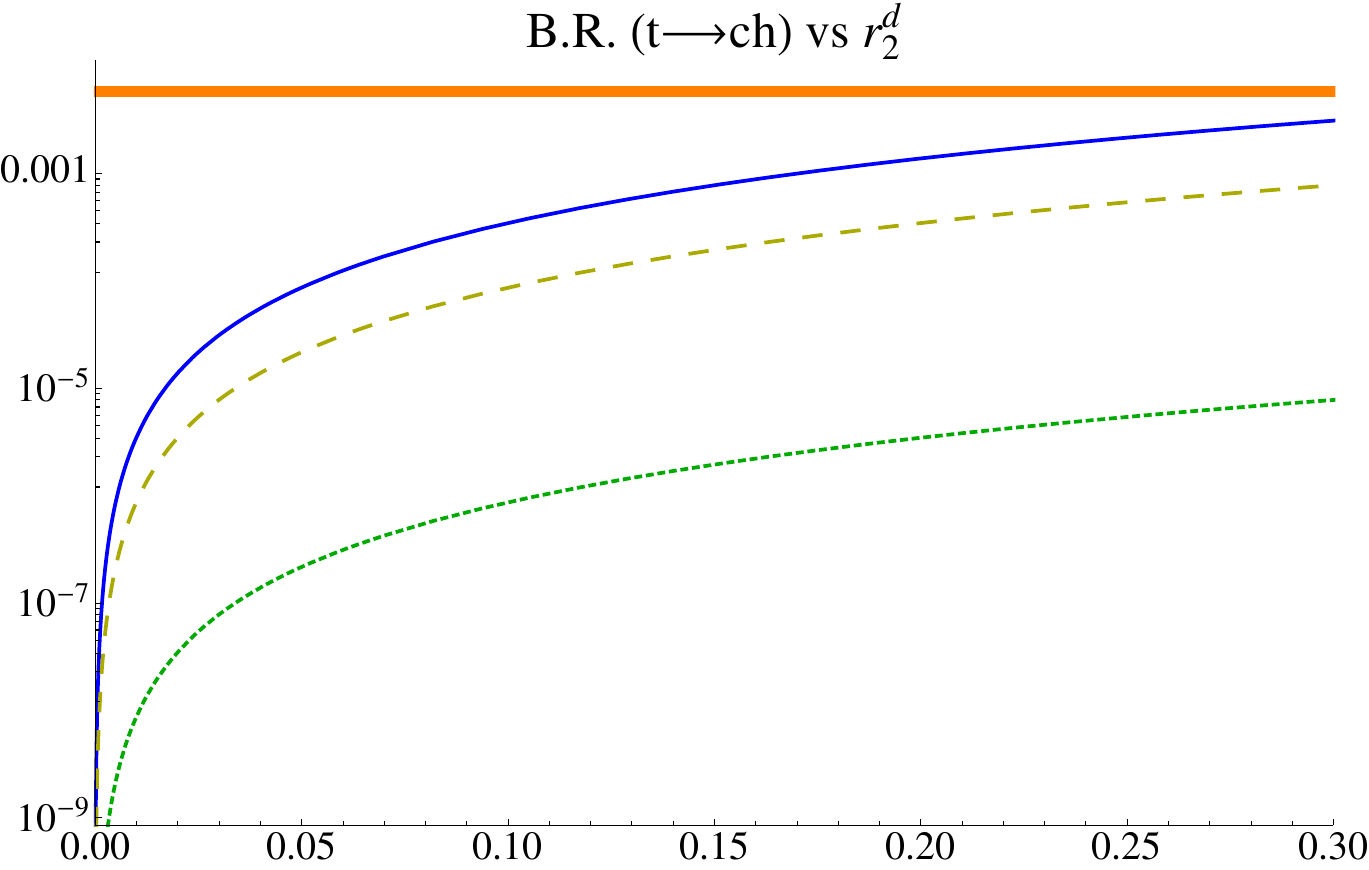}
\caption{ The B.R.($t\rightarrow ch$) dependence to $r_2^d$. The upper bound or horizontal line (orange) comes from the ATLAS and CMS direct constraints \cite{Aad:2014dya,CMS:2014qxa}, whereas the continous line (blue), the large dashing line (yellow), and the small dashing line (green) correspond to the three different taken cases $u=(0.5,1,10$) TeV, respectively. 
\label{Br-plot }}
\end{figure}

Under the approximation that the top decay $t\to b+W$ dominates the total width,
and it is given by the SM result $\Gamma(t\to b+W) \simeq 1.5$ GeV, we obtain:
B.R.$(t\to ch)=0.58 \, |\kappa^u \tilde{Z}_{23}|^2$. 
For $v/u \simeq 0.25$ and the values of the parameters called scenario Z in Table \ref{Zmatrix}, one finds
that the B.R. reaches a value of B.R. $\simeq 3.0\times 10^{-3}$, which is about ten orders of magnitude above
the SM prediction. Moreover, this decay can be tested at the LHC \cite{Greljo:2014dka}.
Values of the B.R. for other choices of parameters, using the  results of Table \ref{Zmatrix} are shown
in Table \ref{Tchbr}. The labels X1, X2, and X3 correspond to the choices of $u=0.5,1,10$ TeV, within scenario X
of Table \ref{Zmatrix}, and similarly for Y1, Y2, and Y3 and  Z1, Z2, and Z3. Figure \ref{Br-plot }
 portraits
the dependence of the B.R.($t\rightarrow ch$) to $r_2^d$ and shows the three different cases $u=(0.5,$ $1$, $10$) TeV. 

\section{Conclusions and outlook}
\label{Sec:VI}

One of the most important tasks of future colliders is to study the properties 
of the  Higgs-like particle with $m_h=125-126$ GeV  discovered at the LHC.
Current measurements of its spin, parity, and interactions, seem consistent with 
the SM, where its couplings to fermions and gauge bosons are proportional to the 
particle mass. We have studied the variations on these couplings, within 
a model with an extended Higgs sector, where the masses for each fermion type,
arise from a different Higgs doublet. These couplings to the Higgs field become $\eta^u=1.03$, $\eta^d = 0.61$, $\eta^\ell=1.04$,
and $\chi_V = 0.96$ under one specific set of values. And their behaviour shows an unexpected feature: d-type quarks should have a 
weaker coupling when compared to the rest, which are in fact above one. And interestingly, recent measurements of the Higgs 
coupling to the different SM particles \cite{Aad:2015gba} report a similar behaviour wherein the bottom quark shows a smaller 
coupling ($\eta^d_\text{exp} = 0.63^{+0.39}_{-0.37}$) while the rest are larger than one. The present precision is too low to 
hint to any strong deviation from the SM as the experimental value easily agrees $(1\sigma)$ with the SM's prediction. 
Nevertheless, if this scenario is confirmed in the years to come it will suggest that the scalar sector has a richer 
structure than the SM one, where one Higgs doublet gives masses to all fermion types.

This model includes mixing of the Higgs 
doublets with a flavon field, which generates the Yukawa hierarchies and induces 
flavor-violating Higgs couplings at acceptable rates.
Constraints on these couplings are derived from the contributions to K and D mixing and from the Higgs searches at the LHC. 
Their implications are discussed. On one hand, from the former set, the following upper bounds to the effective couplings 
are found $\kappa^d \widetilde{z}^d_{12} < 1 \times 10^{-6}$ and $\kappa^u \widetilde{z}^u_{12} < 1 \times 10^{-8}$. 
On the other, from the constraints coming from the Higgs searches at the LHC,
the B.R. of the FCNC top decay $t\to ch$ is considered. We find that this mode could reach
a B.R. of order $10^{-3}$, which could be studied at the LHC. In the down-type quarks and lepton sectors, 
there are also interesting aspects to study in the future, such as the rates for rare b-decays 
or  the decay $h \to \tau\mu$ which could be induced at rates 
that could be detected at future colliders \cite{DiazCruz:1999xe}. The complementarity of future colliders has 
been studied in \cite{Kanemura:2014dea}.

There are two aspects of our model that give some hope to measure the Higgs couplings
with light fermions, namely, DM and LFV. In the case of DM, it is possible that
its interaction with nucleons could be mediated by the Higgs boson, which depends on the strength
of the Higgs interaction with nucleons, which in turn depends on the Higgs coupling with light
fermions. Therefore, by searching for DM-nucleon dispersion, one is also testing the Higgs coupling
with light quarks. Similar remarks hold when one considers $e-\mu$ conversion, where the
Higgs nucleon interaction appears. Thus,  nature could be extra benevolent, and  besides 
showing evidence of the new physics, it would allow to test the couplings
of light quarks with the Higgs.

However, the most salient feature  of this model is the flavon field, which has FV couplings
with fermions. Within the scenarios discussed here, the fourth Higgs boson $H_F$ is
dominated by its flavon ($s_1$) component. The size of its coupling with top quarks, 
will be given by the factor $\kappa_4 \,\tilde{Z}^u_{33}$, where:
$ \kappa_4 = O^T_{14} \cos\beta_1 v/u$, which is of order $0.1$ for  $v/u=0.1$,
while  $\tilde{Z}^u_{33}\simeq 0.5$ (within the scenario that we called Z appearing in Table \ref{Zmatrix}).
Thus, we will have a coupling of order $10^{-2}$ (compare with the SM top-Higgs
coupling which is of order 0.7). Thus $H_F$ could be produced through gluon fusion (top loop), 
with a cross-section of order $3\times 10^{-4}$ times the SM Higgs cross section. 
 Despite the fact that the cross section seems quite suppressed, one has the advantage of 
the clean signature of flavon decays, such as the LFV mode $H_F \to \tau \mu$,
which will have a B.R. similar to the ones for the FC modes.
 A detailed study of this mechanism at the very Large Hadron Collider with 100 TeV c.m. energy
 is underway \cite{OurFlvnVLHC}.

\begin{acknowledgments}
We would like to acknowledge useful discussions with
E. D\'iaz, O. Felix-Beltr\'an, F. Gonz\'alez-Canales, B. Larios, and R. Noriega-Papaqui.
This work has been supported by CONACyT-Mexico under
Contract No.~220498, CONACYT-SNI (Mexico), and VIEP (BUAP).
\end{acknowledgments}

\appendix

\section{The scalar potential for the 3 + 1 + FN model}
\label{ScalarPotential}

The scalar sector of the model is made out of five scalar fields from which  four of them are doublets of $SU(2)_L$, $\Phi_i$ ($i=1,2,3,4$) while the remaining one, $S$, is a singlet. From the former four one of them, $\Phi_4$, is inert, giving us a viable candidate for Dark Matter (DM). The other three, namely,
$\Phi_1$, $\Phi_2$, and $\Phi_3$ are coupling each of them to all the fermions of the same electric charge, i.e., to up- and down-type quarks and charged leptons, respectively. This scenario would naturally avoid FCNC's if the singlet $S$, which serves as a Froggatt Nielsen (FN) or flavon field, would have not been introduced. The singlet $S$ allows, through the addition of the FN mechanism, a straightforward description of the hierarchical structure in the Yukawa matrices which in the end gives the observed pattern of hierarchical fermion masses, that is, $m_3 \gg m_2 \gg m_1$. 

Now, if explicitly written in terms of its components the scalar content is 
\begin{eqnarray}
	\Phi_i = \begin{pmatrix}
	\varphi_i^+ \\
	\frac{v_i + \phi_i^0 + i \chi^0_i}{\sqrt{2}}
	\end{pmatrix}, \; (i=1,2,3) \qquad
	\Phi_4 = \begin{pmatrix}
	s^+ \\
	\frac{s^0 + i P^0}{\sqrt{2}}
	\end{pmatrix},
\end{eqnarray}
and
\begin{eqnarray}
	S = \frac{1}{\sqrt{2}}(u + s_1 + i s_2).
\end{eqnarray}
Notice that $s^0$ or $P^0$ in the inert scalar, $\Phi_4$, could be DM. 

Then, the potential is
\begin{eqnarray}
	V = V_{3H} + V_N + V_S + V_{HN} + V_{SH},
\end{eqnarray}
where
\begin{eqnarray}
	V_{3H} = -\frac{1}{2} \left[ m_{11}^2 \Phi_1^\dagger \Phi_1 + m_{22}^2 \Phi_2^\dagger \Phi_2 + m_{33}^2 \Phi_3^\dagger \Phi_3 \right] +
	\left[ m_{12}^2 \Phi_1^\dagger \Phi_2 + m_{13}^2 \Phi_1^\dagger \Phi_3 + m_{23}^2 \Phi_2^\dagger \Phi_3 + h.c. \right] \\ \nonumber
	+ \left[ \frac{\lambda_{11}}{2} A_{1}^2 + \frac{\lambda_{22}}{2} A_{2}^2 +  \frac{\lambda_{33}}{2} A_{3}^2 +  \lambda_{12} \lambda_{12} A_1 A_2 + \lambda_{13} A_1 A_3 + \lambda_{23} A_2 A_3 \right] \\ \nonumber
	+ \left[ \lambda_{412} A_{12}^\dagger A_{12} + \lambda_{413} A_{13}^\dagger A_{13}
	+\lambda_{423} A_{23}^\dagger A_{23} \right] 
	+ \left[ \lambda_{512} (A_{12})^2 + \lambda_{513} (A_{13})^2 + \lambda_{523} (A_{23})^2 + h.c. \right] ,
	\\
	V_{N} = \frac{1}{2} m_{44}^2 \Phi^\dagger_4 \Phi_4 + \frac{\lambda_{44}}{2} (\Phi^\dagger_4 \Phi_4)^2, \\
	V_S =  m_S^2 S^\dagger S + \lambda_S (S^\dagger S)^2 + \widetilde{m}_S^2 (S^2 + S^*{}^2),\\
	V_{HN} = \left[ \widetilde{\lambda}_{11} \Phi_1^\dagger \Phi_1 + \widetilde{\lambda}_{22} \Phi_2^\dagger \Phi_2 + \widetilde{\lambda}_{33} \Phi_3^\dagger \Phi_3 \right] (\Phi_4^\dagger \Phi_4) + \widetilde{\lambda}_{14} A_{14}^\dagger A_{14} + \widetilde{\lambda}_{24} A_{24}^\dagger A_{24} + \widetilde{\lambda}_{34} A_{34}^\dagger A_{34} \\ \nonumber
	+ \frac{1}{2} \left[ \lambda'_{14} A_{14}^2 + \lambda'_{24} A_{24}^2 + \lambda'_{34} A_{34}^2 + h.c. \right], \\
	V_{HS} = \left[ \lambda_{S1} A_1 + \lambda_{S2} A_2 + \lambda_{S3} A_3\right] (S^* S),
\end{eqnarray}
and where we have denoted by $A_i = \Phi_i^\dagger \Phi_i$ and $A_{jk} = \Phi_j^\dagger \Phi_k$.


\begin{thebibliography}{99}

\bibitem{higgs-atlas:2012gk}
  G.~Aad {\it et al.}  [ATLAS Collaboration],
  Phys.\ Lett.\ B {\bf 716}, 1 (2012)
  [arXiv:1207.7214 [hep-ex]].


\bibitem{higgs-cms:2012gu}
  S.~Chatrchyan {\it et al.}  [CMS Collaboration],
  Phys.\ Lett.\ B {\bf 716}, 30 (2012)
  [arXiv:1207.7235 [hep-ex]].

\bibitem{Gunion:1989we}
  J.~F.~Gunion, H.~E.~Haber, G.~L.~Kane and S.~Dawson,
  Front.\ Phys.\  {\bf 80}, 1 (2000).
  

\bibitem{Erler:2007sc}
  J.~Erler,
  AIP Conf.\ Proc.\  {\bf 917}, 244 (2007)
  [hep-ph/0701261].

\bibitem{Pomarol:2012sb}
  A.~Pomarol,
  CERN Yellow Report CERN-2012-001, 115-151
  [arXiv:1202.1391 [hep-ph]].


\bibitem{Martin:1997ns}
  S.~P.~Martin,
  In *Kane, G.L. (ed.): Perspectives on supersymmetry II* 1-153
  [hep-ph/9709356].
  
  
\bibitem{Kane:2006hd}
  G.~L.~Kane, P.~Kumar, D.~E.~Morrissey and M.~Toharia,
  Phys.\ Rev.\ D {\bf 75}, 115018 (2007)
  [hep-ph/0612287].

\bibitem{Arganda:2013ve}
  E.~Arganda, J.~L.~Diaz-Cruz and A.~Szynkman,
  Phys.\ Lett.\ B {\bf 722}, 100 (2013)
  [Phys.\ Lett.\ B {\bf 722}, 100 (2013)]
  [arXiv:1301.0708 [hep-ph]].

\bibitem{Arganda:2012qp}
  E.~Arganda, J.~L.~Diaz-Cruz and A.~Szynkman,
  Eur.\ Phys.\ J.\ C {\bf 73}, 2384 (2013)
  [arXiv:1211.0163 [hep-ph]].

\bibitem{Chakraborty:2013si}
  A.~Chakraborty, B.~Das, J.~L.~Diaz-Cruz, D.~K.~Ghosh, S.~Moretti and P.~Poulose,
  arXiv:1301.2745 [hep-ph].
  
\bibitem{Botella:2016krk} 
  F.~J.~Botella, G.~C.~Branco, M.~N.~Rebelo and J.~I.~Silva-Marcos,
  arXiv:1602.08011 [hep-ph].
  
\bibitem{DiazCruz:2007be} 
  J.~L.~Diaz-Cruz,
  Phys.\ Rev.\ Lett.\  {\bf 100}, 221802 (2008)
  doi:10.1103/PhysRevLett.100.221802
  [arXiv:0711.0488 [hep-ph]].
  
\bibitem{Veltman:1980mj} 
  M.~J.~G.~Veltman,
  Acta Phys.\ Polon.\ B {\bf 12}, 437 (1981).


\bibitem{Grossman:1994jb} 
  Y.~Grossman,
  Nucl.\ Phys.\ B {\bf 426}, 355 (1994)
  doi:10.1016/0550-3213(94)90316-6
  [hep-ph/9401311].

\bibitem{Branco:2011iw}
  G.~C.~Branco, P.~M.~Ferreira, L.~Lavoura, M.~N.~Rebelo, M.~Sher and J.~P.~Silva,
  Phys.\ Rept.\  {\bf 516}, 1 (2012)
  [arXiv:1106.0034 [hep-ph]].

\bibitem{DiazCruz:2010yq}
  J.~L.~Diaz-Cruz, A.~Diaz-Furlong and J.~H.~Montes de Oca,
  arXiv:1010.0950 [hep-ph].
  
  
\bibitem{DiazCruz:2004tr}
   J.~L.~Diaz-Cruz, R.~Noriega-Papaqui and A.~Rosado,
  Phys.\ Rev.\ D {\bf 69}, 095002 (2004)
  [hep-ph/0401194].
  
    
\bibitem{Cotti:2002zq} 
  U.~Cotti, J.~L.~Diaz-Cruz, R.~Gaitan, H.~Gonzales and A.~Hernandez-Galeana,
  Phys.\ Rev.\ D {\bf 66}, 015004 (2002)
  [hep-ph/0205170].

\bibitem{DiazCruz:2002er}
  J.~L.~Diaz-Cruz,
  JHEP {\bf 0305}, 036 (2003)
  [hep-ph/0207030];
  
  

\bibitem{Ellis:2014dva} 
  J.~Ellis, V.~Sanz and T.~You,
  arXiv:1404.3667 [hep-ph].
  

\bibitem{Espinosa:2012ir} 
  J.~R.~Espinosa, C.~Grojean, M.~Muhlleitner and M.~Trott,
  JHEP {\bf 1205}, 097 (2012)
  [arXiv:1202.3697 [hep-ph]].
 
\bibitem{Klute:2012pu} 
  M.~Klute, R.~Lafaye, T.~Plehn, M.~Rauch and D.~Zerwas,
  Phys.\ Rev.\ Lett.\  {\bf 109}, 101801 (2012)
  [arXiv:1205.2699 [hep-ph]].
  
\bibitem{Grinstein:2013fia} 
  B.~Grinstein, C.~W.~Murphy, D.~Pirtskhalava and P.~Uttayarat,
  JHEP {\bf 1405}, 083 (2014)
  [arXiv:1401.0070 [hep-ph]].
  
\bibitem{Espinosa:2012im} 
  J.~R.~Espinosa, C.~Grojean, M.~Muhlleitner and M.~Trott,
  JHEP {\bf 1212}, 045 (2012)
  [arXiv:1207.1717 [hep-ph]].
 
\bibitem{Cheung:2013rva} 
  K.~Cheung, J.~S.~Lee and P.~-Y.~Tseng,
  JHEP {\bf 1401}, 085 (2014)
  [arXiv:1310.3937 [hep-ph]].
  
    \bibitem{Fritzsch:1999ee}
H.~Fritzsch and Z.-z. Xing, ``{Mass and flavor mixing schemes of quarks and
  leptons},'' \href{http://dx.doi.org/10.1016/S0146-6410(00)00102-2}{{\em
  Prog.Part.Nucl.Phys.} {\bfseries 45} (2000) 1--81},
\href{http://arxiv.org/abs/hep-ph/9912358}{{\ttfamily arXiv:hep-ph/9912358
  [hep-ph]}}.

  \bibitem{King:2013eh}
S.~F. King and C.~Luhn, ``{Neutrino Mass and Mixing with Discrete Symmetry},''
  \href{http://dx.doi.org/10.1088/0034-4885/76/5/056201}{{\em Rept.Prog.Phys.}
  {\bfseries 76} (2013) 056201},
\href{http://arxiv.org/abs/1301.1340}{{\ttfamily arXiv:1301.1340 [hep-ph]}}.

\bibitem{Hollik:2014jda}
W.~G. Hollik and U.~J. Saldana-Salazar.
\newblock {The double mass hierarchy pattern: simultaneously understanding
  quark and lepton mixing}.
\newblock {\em Nucl.Phys.}, B892:364--389, 2015.
 
  
\bibitem{Glashow:1976nt} 
  S.~L.~Glashow and S.~Weinberg,
  Phys.\ Rev.\ D {\bf 15}, 1958 (1977).
  doi:10.1103/PhysRevD.15.1958
  
  
\bibitem{Ma:2006fn}
  E.~Ma,
  Mod.\ Phys.\ Lett.\ A {\bf 21}, 1777 (2006)
  [hep-ph/0605180].
  
\bibitem{Grzadkowski:2010au} 
  B.~Grzadkowski, O.~M.~Ogreid, P.~Osland, A.~Pukhov and M.~Purmohammadi,
  JHEP {\bf 1106}, 003 (2011)
  [arXiv:1012.4680 [hep-ph]].
   
   
\bibitem{Krawczyk:2013jta} 
  M.~Krawczyk, D.~Sokolowska, P.~Swaczyna and B.~Swiezewska,
  JHEP {\bf 1309}, 055 (2013)
  [arXiv:1305.6266 [hep-ph]].
   
  
\bibitem{Dorsner:2002wi} 
  I.~Dorsner and S.~M.~Barr,
  Phys.\ Rev.\ D {\bf 65}, 095004 (2002)
  [hep-ph/0201207].
  
  
\bibitem{Tsumura:2009yf} 
  K.~Tsumura and L.~Velasco-Sevilla,
  Phys.\ Rev.\ D {\bf 81}, 036012 (2010)
  [arXiv:0911.2149 [hep-ph]].
   

\bibitem{Berger:2014gga} 
  E.~L.~Berger, S.~B.~Giddings, H.~Wang and H.~Zhang,
  Phys.\ Rev.\ D {\bf 90}, no. 7, 076004 (2014)
  [arXiv:1406.6054 [hep-ph]].
  
\bibitem{Hirsch:2010ru} 
  M.~Hirsch, S.~Morisi, E.~Peinado and J.~W.~F.~Valle,
  Phys.\ Rev.\ D {\bf 82}, 116003 (2010)
  doi:10.1103/PhysRevD.82.116003
  [arXiv:1007.0871 [hep-ph]].
  
\bibitem{Bolanos:2016aik} 
  A.~Bola\~nos, J.~L.~Diaz-Cruz, G.~Hern\'andez-Tom\'e and G.~Tavares-Velasco,
  arXiv:1604.04822 [hep-ph].
  
\bibitem{Bonilla:2014xba}
  C.~Bonilla, D.~Sokolowska, N.~Darvishi, J.~L.~Diaz-Cruz and M.~Krawczyk,
  J.\ Phys.\ G {\bf 43} (2016) no.6,  065001
  doi:10.1088/0954-3899/43/6/065001
  [arXiv:1412.8730 [hep-ph]].
  
\bibitem{DiazCruz:1992uw} 
  J.~L.~Diaz-Cruz and A.~Mendez,
  Nucl.\ Phys.\ B {\bf 380}, 39 (1992).
  doi:10.1016/0550-3213(92)90514-C
  
\bibitem{Barroso:2006pa} 
  A.~Barroso, P.~M.~Ferreira, R.~Santos and J.~P.~Silva,
  Phys.\ Rev.\ D {\bf 74}, 085016 (2006)
  doi:10.1103/PhysRevD.74.085016
  [hep-ph/0608282].
  
\bibitem{Barroso:2005sm} 
  A.~Barroso, P.~M.~Ferreira and R.~Santos,
  Phys.\ Lett.\ B {\bf 632}, 684 (2006)
  doi:10.1016/j.physletb.2005.11.031
  [hep-ph/0507224].
  
\bibitem{Ferreira:2004yd} 
  P.~M.~Ferreira, R.~Santos and A.~Barroso,
  Phys.\ Lett.\ B {\bf 603}, 219 (2004)
  Erratum: [Phys.\ Lett.\ B {\bf 629}, 114 (2005)]
  doi:10.1016/j.physletb.2004.10.022, 10.1016/j.physletb.2005.09.074
  [hep-ph/0406231].
  
    
\bibitem{Ivanov:2012fp} 
  I.~P.~Ivanov and E.~Vdovin,
  Eur.\ Phys.\ J.\ C {\bf 73}, 2309 (2013)
  [arXiv:1210.6553 [hep-ph]].

\bibitem{Keus:2013hya} 
  V.~Keus, S.~F.~King and S.~Moretti,
  JHEP {\bf 1401}, 052 (2014)
  [arXiv:1310.8253 [hep-ph]].

\bibitem{Arkani-Hamed:2015vfh} 
  N.~Arkani-Hamed, T.~Han, M.~Mangano and L.~T.~Wang,
  arXiv:1511.06495 [hep-ph].
  
\bibitem{Dery:2014kxa} 
  A.~Dery, A.~Efrati, Y.~Nir, Y.~Soreq and V.~Susič,
  Phys.\ Rev.\ D {\bf 90}, 115022 (2014)
  doi:10.1103/PhysRevD.90.115022
  [arXiv:1408.1371 [hep-ph]].
  
\bibitem{Huitu:2016pwk} 
  K.~Huitu, V.~Keus, N.~Koivunen and O.~Lebedev,
  JHEP {\bf 1605}, 026 (2016)
  doi:10.1007/JHEP05(2016)026
  [arXiv:1603.06614 [hep-ph]].
  
\bibitem{Dreiner:2003yr} 
  H.~K.~Dreiner, H.~Murayama and M.~Thormeier,
  Nucl.\ Phys.\ B {\bf 729}, 278 (2005)
  doi:10.1016/j.nuclphysb.2005.08.047
  [hep-ph/0312012].
  
\bibitem{Ferreira:2014naa} 
  P.~M.~Ferreira, J.~F.~Gunion, H.~E.~Haber and R.~Santos,
  arXiv:1403.4736 [hep-ph].
 
  
\bibitem{Giardino:2013bma} 
  P.~P.~Giardino, K.~Kannike, I.~Masina, M.~Raidal and A.~Strumia,
  arXiv:1303.3570 [hep-ph].
  
\bibitem{Aad:2015gba} 
  G.~Aad {\it et al.} [ATLAS Collaboration],
  Eur.\ Phys.\ J.\ C {\bf 76}, no. 1, 6 (2016)
  doi:10.1140/epjc/s10052-015-3769-y
  [arXiv:1507.04548 [hep-ex]].
  
\bibitem{Bishara:2016jga} 
  F.~Bishara, U.~Haisch, P.~F.~Monni and E.~Re,
  arXiv:1606.09253 [hep-ph].
  
\bibitem{Soreq:2016rae} 
  Y.~Soreq, H.~X.~Zhu and J.~Zupan,
  arXiv:1606.09621 [hep-ph].
    
\bibitem{Balazs:1998nt} 
  C.~Balazs, J.~L.~Diaz-Cruz, H.~J.~He, T.~M.~P.~Tait and C.~P.~Yuan,
  Phys.\ Rev.\ D {\bf 59}, 055016 (1999)
  [hep-ph/9807349].
  
\bibitem{DiazCruz:1998qc} 
  J.~L.~Diaz-Cruz, H.~-J.~He, T.~M.~P.~Tait and C.~P.~Yuan,
  Phys.\ Rev.\ Lett.\  {\bf 80}, 4641 (1998)
  [hep-ph/9802294].
  
  
\bibitem{Buras:2010mh}
  A.~J.~Buras, M.~V.~Carlucci, S.~Gori and G.~Isidori,
  JHEP {\bf 1010}, 009 (2010)
  [arXiv:1005.5310 [hep-ph]].

\bibitem{Crivellin:2013wna}
  A.~Crivellin, A.~Kokulu and C.~Greub,
  Phys.\ Rev.\ D {\bf 87}, no. 9, 094031 (2013)
  [arXiv:1303.5877 [hep-ph]].
  
\bibitem{Nierste:2009wg} 
  U.~Nierste,
  arXiv:0904.1869 [hep-ph].
  
\bibitem{Blum:2009sk} 
  K.~Blum, Y.~Grossman, Y.~Nir and G.~Perez,
  Phys.\ Rev.\ Lett.\  {\bf 102}, 211802 (2009)
  doi:10.1103/PhysRevLett.102.211802
  [arXiv:0903.2118 [hep-ph]].
  
\bibitem{Agashe:2014kda} 
  K.~A.~Olive {\it et al.} [Particle Data Group Collaboration],
  Chin.\ Phys.\ C {\bf 38}, 090001 (2014).
  doi:10.1088/1674-1137/38/9/090001

\bibitem{Branco:1999fs} 
  G.~C.~Branco, L.~Lavoura and J.~P.~Silva,
  Int.\ Ser.\ Monogr.\ Phys.\  {\bf 103}, 1 (1999).

\bibitem{Barberio:2008fa} 
  E.~Barberio {\it et al.} [Heavy Flavor Averaging Group Collaboration],
  arXiv:0808.1297 [hep-ex].
  
  \bibitem{LDCtopfv}
  J.~L.~D\'iaz-Cruz, R.~Martinez, M.~A.~Perez and A.~Rosado,
  Phys.\ Rev.\ D {\bf 41}, 891 (1990).
  
\bibitem{Eilam:1991}
G. Eilam, J. L. Hewett and A. Soni, Phys. Rev. D \textbf{44}, 1473 (1991) [Erratum-ibid. D
\textbf{59}, 039901 (1999)].

\bibitem{Mele:1998}
B. Mele, S. Petrarca, and A. Soddu, Phys. Lett. B \textbf{435}, 401 (1998)
{[hep-ph/9805498]}. 



\bibitem{Aguilar:2003}
J. A. Aguilar-Saavedra and B. M. Nobre, Phys. Lett. B \textbf{553}, 251 (2003)
{[hep-ph/0210360]}.


\bibitem{Jenkins:1996zd} 
  E.~E.~Jenkins,
  Phys.\ Rev.\ D {\bf 56}, 458 (1997)
  [hep-ph/9612211].

\bibitem{Bar-Shalom:1998}
S. Bar-Shalom, G. Eilam, A. Soni, and J. Wudka, 
Phys. Rev. Lett. \textbf{79}, 1217 (1997) 
{[hep-ph/9703221]}; 
Phys. Rev. D \textbf{57}, 2957 (1998) 
{[hep-ph/9708358]}.


\bibitem{Bar-Shalom:2005}
S. Bar-Shalom, G. Eilam, M. Frank, and I. Turan, 
Phys. Rev. D \textbf{72}, 055018 (2005) 
{[hep-ph/0506167]}. 


\bibitem{DiazCruz:1999ab} 
  J.~L.~Diaz-Cruz, M.~A.~Perez, G.~Tavares-Velasco and J.~J.~Toscano,
  Phys.\ Rev.\ D {\bf 60}, 115014 (1999)
  [hep-ph/9903299].
  

\bibitem{Frank:2006}
M. Frank and I. Turan, Phys. Rev. D \textbf{74}, 073014 (2006)
{[hep-ph/0609069]};

\bibitem{DiazCruz:2012xa} 
  J.~L.~Diaz-Cruz, A.~Diaz-Furlong, R.~Gaitan-Lozano and J.~H.~Montes de Oca Y.,
  Eur.\ Phys.\ J.\ C {\bf 72}, 2119 (2012)
  [arXiv:1203.6889 [hep-ph]].




\bibitem{Larios:2006}
F. Larios, R. Mart\'inez, and M. A. P\'erez,
Int. J. Mod. Phys. A \textbf{21}, 3437 (2006)
{[hep-ph/0605003]}.

\bibitem{Quintero:2014lqa} 
  N.~Quintero, J.~L.~Diaz-Cruz and G.~Lopez Castro,
  arXiv:1403.3044 [hep-ph].

\bibitem{Greljo:2014dka} 
  A.~Greljo, J.~F.~Kamenik and J.~Kopp,
  arXiv:1404.1278 [hep-ph].
  
\bibitem{Aad:2014dya} 
  G.~Aad {\it et al.} [ATLAS Collaboration],
  JHEP {\bf 1406}, 008 (2014)
  doi:10.1007/JHEP06(2014)008
  [arXiv:1403.6293 [hep-ex]].
  
\bibitem{CMS:2014qxa} 
  CMS Collaboration [CMS Collaboration],
  CMS-PAS-HIG-13-034.
  
\bibitem{DiazCruz:1999xe} 
  J.~L.~Diaz-Cruz and J.~J.~Toscano,
  Phys.\ Rev.\ D {\bf 62}, 116005 (2000)
  [hep-ph/9910233].
  
\bibitem{Kanemura:2014dea} 
  S.~Kanemura, H.~Yokoya and Y.~-J.~Zheng,
  arXiv:1404.5835 [hep-ph].


\bibitem{OurFlvnVLHC}
J.L. Diaz-Cruz et al.,work in progress.



  
  

    
  


  
  

\end{thebibliography}
\end{document}